\documentclass [aps,12pt,aps,a4paper,prb]{revtex4}
\usepackage{graphicx}  
\usepackage{dcolumn}   
\usepackage{bm}        
\usepackage{amssymb}   
\usepackage{natbib}

\begin{document}



\title{Theory of polyzwitterion conformations}
\author{{\bf Rajeev Kumar and Glenn H. Fredrickson} \footnote[1]{To whom any correspondence should be addressed, Email : ghf@mrl.ucsb.edu}}

\affiliation{Materials Research Laboratory, University of California, Santa Barbara, CA-93106-5080}
\date{\today}

\begin{abstract}
Conformational characteristics of polyzwitterionic molecules in aqueous solutions are investigated using the variational
method. Analytical relations are derived for the radius of gyration of a single polyzwitterionic chain as a function of the chain length, electrostatic interaction strength, added salt concentration, dipole moment and degree of ionization of the zwitterionic monomers. In the absence of the small ions (counterions and coions) 
near the polyzwitterionic chain, 
attractive dipole-dipole interactions are shown to induce a collapse of the polyzwitterionic chain. However, in the presence of the small ions, the radius of gyration is shown to be an interplay of the screening of the electrostatic interactions and the counterion adsorption on the zwitterionic sites. In addition to the well-known 
Debye-H\"{u}ckel screening of the charge-charge interactions, screening of the charge-dipole and dipole-dipole interactions are found to play important roles in determining the size of the chain. Functional
forms for the screened charge-dipole and dipole-dipole interaction potentials are presented. Furthermore, counterion
adsorption on the zwitterionic monomers is predicted to be asymmetric depending on the nature of the added salt and the zwitterionic groups. Qualitative remarks regarding the  solubility of these molecules in aqueous solutions along with the classical ``anti-polyelectrolyte'' effect (increase in the solubility in water with the addition of salt) are presented. 
\end{abstract}

\maketitle

\section{Introduction}
Polyzwitterions\cite{morawetz57,timmerman58,mccormick_review02}
 belong to a special class of polyampholytes\cite{rubinstein_review04,kudaibergenov_review06}, where each monomer carries both the positive and negative charges. Different kinds of polyzwitterions\cite{mccormick_review02,rubinstein_review04,kudaibergenov_review06} have
been synthesized by varying the length of the zwitterionic unit,
the spacing and the functionality of the group attaching the zwitterionic unit to the chain backbone.
Also, depending on the synthesis scheme, polyzwitterions can be prepared without any counterions of the
charged moieties on the zwitterionic unit or with counterions. In the literature, 
a major class of such molecules are known as betaines\cite{mccormick_review02,rubinstein_review04,kudaibergenov_review06} 
and, depending on the functionality of the negatively charged site on the zwitterionic unit, 
are called sulfo- (i.e., sulphonate group in the zwitterion unit), carbo- (carboxylate group) and phospho- (phosphate) betaines. These molecules have a wide variety of applications\cite{kudaibergenov_review06}
in medicine, biotechnology and oil industry.

Since the synthesis of first polyzwitterionic molecule in 1957\cite{morawetz57}, a number of experimental studies\cite{kudaibergenov_review06,salamone78,schulz86,wu00,bendejacq07} have been carried out on this special
class of polyampholytes. One of the well-known signatures of these molecules is their insolubility in water 
despite the presence of charged groups. Furthermore, it is observed that the addition of salt  
enhances the solubility of these molecules. This particular effect is known as ``anti-polyelectrolyte''
effect\cite{mccormick_review02,bendejacq07} in the literature due to the opposite behavior seen in pure 
polyelectrolytes\cite{muthubeer97,joanny04}. Furthermore, the enhancement in the solubility of
these molecules in water has been shown to depend on the specific nature of the 
added salt\cite{salamone78,schulz86,bendejacq07}. For polycarbobetaines, the solubility
also depends on pH of the solution, due to the presence of an acid-base equilibrium mechanism in
these systems. 

In the literature\cite{kudaibergenov_review06,salamone78,schulz86,wu00,bendejacq07}, insolubility of 
polyzwitterions in pure water is explained on intutive grounds by the presence of attractive dipolar intra and 
inter-molecular interactions in salt-free conditions. Despite the lack of screening in purely dipolar 
media\cite{park79}, the screening of attractive dipolar interactions by the added salt\cite{phillies74} 
is conjectured to be responsible for the ``anti-polyelectrolyte'' effect. The minimum amount of salt required to 
solubilize the polyzwitterions\cite{kudaibergenov_review06,salamone78} is further observed to depend on 
specific interactions between the charged
groups on the zwitterionic unit and the salt ions. Sometimes\cite{bendejacq07}, both the ``polyelectrolyte'' 
(decrease in the solubility of these molecules in aqueous solutions with the addition of salt) 
and ``anti-polyelectrolyte'' effects are observed in these systems. Using zeta-potentiometry\cite{bendejacq07}, 
it has been shown that the ``polyelectrolyte'' 
effect is a consequence of a net charge on the polyzwitterionic chain.  
The role of asymmetric counterion adsorption on the solubility of polyzwitterions has also 
been revealed using zeta-potentiometry. 

On the theoretical front, a number of studies have been carried out on polyampholyte
solutions where the molecules contain both positive and negative charges \textit{along} the backbone, 
starting from the work by Edwards-King-Pincus\cite{ekp80}. The conformations of polyampholytes were 
investigated in detail by Higgs and Joanny\cite{higgs91}, Kantor and Kardar\cite{kardar92}
and Dobrynin and Rubinstein\cite{rubinstein95}. It has been shown that the overall net charge\cite{higgs91,rubinstein95} 
of the chain, arising as a result of differential counterion adsorption on the positive and negative charges, 
plays a crucial role. Furthermore, it has been predicted that depending on the extent of asymmetry in the counterion 
adsorption on the charged sites along the backbone, the polyampholyte chain may behave 
like a polyelectrolyte. Also, the charge sequence along the polyampholyte chain\cite{srivastava96} and the formation 
of ionic bridges\cite{khokhlov01} between the oppositely charged groups along the backbone have been 
shown to affect the conformational behavior. In contrast to these polyampholyte models, polyzwitterions bear 
zwitterionic side-groups carrying both positive 
and negative charges on the same monomeric unit. The presence of the positive and negative charges on the same 
monomer implies that dipolar interactions\cite{intermolecular_forces,sluckin94,muthu96,podgornik04} 
are of paramount importance in these systems, which complicates the analysis. So far, polyzwitterions 
have received little attention from the theoretical community.

In this work, we consider a single polyzwitterionic chain and derive quantitative relations between 
the radius of gyration of the chain and the added salt concentration using a variational 
technique\cite{edwardsbook} that was previously applied to neutral polymers\cite{edwards_singh} and 
polyelectrolytes\cite{muthu87,muthu04}. With the aid of these 
relations, experimental observations regarding the solubility of polyzwitterions can be inferred.  
Here, an increase or decrease in the solubility of polymers on changing experimental conditions is inferred from 
the increase or decrease in the radius of gyration of the chain, respectively. 
Using the radius of gyration as a gauge for measuring the solubility of polymers, concepts of poor, theta and good solvent 
conditions have already been developed for neutral polymers\cite{scalingbook}. In this work, we carry out a similar 
analysis for a polyzwitterionic chain and study the effect of added salt on the radius of gyration of the chain, 
which, in turn, 
is used to infer the effect of added salt on the solubility of polyzwitterionic molecules.  

Before presenting the theoretical outline of this paper, we summarize the 
assumptions and key results of the model for those who are uninterested in the mathematical details. 

(i) We model the polyzwitterionic chain of $N$ Kuhn segments by a continuous curve of 
length $Nl$, $l$ being the length of each segment, where each segment has a dipole of length $r_d$ attached to it at an angle. We consider the situation in which the chain is surrounded by two kinds (positive and negative) of monovalent small ions  arising from the dissociation of the zwitterionic monomers along with those coming from the added monovalent salt. We assume that there are $\alpha_+ N$ positive and $\alpha_- N$ negative counterions released by the zwitterionic monomers. Furthermore, we assume that the negative ($= -e \alpha_+ N, e$ being the electronic charge) and positive charges ($= e\alpha_- N$) are distributed uniformly among the negative and positive sites, respectively. Also, the interactions between the segments with electric dipoles are modeled by a short range delta function pseudopotential of strength $w$ and the long range electrostatic  interactions between the charges of the electric dipoles. Dividing the population of the small ions into ones free to move in the solution (called ``free'' ions) and those, which are adsorbed on the zwitterionic monomers (``adsorbed'' ions), we compute the radius of gyration of the polyzwitterionic chain using a variational method in case the electric dipoles can rotate freely.

(ii) The counterion adsorption in the presence of the small ions near the polyzwitterionic chain complicates the situation. This is because the counterion adsorption leads to the formation of small dipoles at the adsorption sites. So, the description of the segment-segment interactions must include the  charge-charge and charge-dipole interactions in addition to the dipole-dipole interactions. Integrating over the positions of the ``free'' ions, it is found that in addition to the well-known Debye-H\"{u}ckel screening of the charge-charge interactions, charge-dipole and dipole-dipole interactions also get screened. For a charge of magnitude $Q$ (in units of electronic charge) and a dipole of moment $\mathbf{p}$ (in units of electronic charge) located at $\mathbf{r}$ and $\mathbf{r}'$, respectively, charge-dipole interaction energy (in units of $k_B T, k_B$ being the Boltzmann constant and $T$ being the temperature) is found to be 
\begin{eqnarray}
W_{cd}(\mathbf{r},Q,\mathbf{r}',\mathbf{p})  &=& -l_B Q \frac{e^{-\kappa |\mathbf{r}-\mathbf{r}'|}}{|\mathbf{r}-\mathbf{r}'|^3} \left[1 + \kappa |\mathbf{r}-\mathbf{r}'| \right ]\left[\mathbf{p}.(\mathbf{r}-\mathbf{r}')\right],
\end{eqnarray}
where $l_B = e^{2}/4\pi \epsilon_0 \epsilon_r k_B T$ is Bjerrum's length written in terms of the relative permittivity $\epsilon_r$ of the medium, $\epsilon_0$ being the permittivity of vacumm. Also, $1/\kappa$ is Debye length having contributions from 
only the ``free'' small ions in the solution given by 
\begin{eqnarray}
\kappa^2 &=& 4\pi l_B \sum_{j} Z_j^2 n_j/\Omega,
\end{eqnarray}
where $j=c+,c-,s+,s-$ represents the positive, negative ions dissociated from the zwitterionic monomers and the added salt. For the charge and a freely rotating dipole, the interaction energy depends on the magnitude of the dipole moment ($p = |\mathbf{p}|$) and is given by  
\begin{eqnarray}
W_{cd}^{f}(\mathbf{r},Q,\mathbf{r}',\mathbf{p})  &=& -\frac{l_B^2}{6} Q^2 p^2 \frac{e^{-2\kappa |\mathbf{r}-\mathbf{r}'|}}{|\mathbf{r}-\mathbf{r}'|^4} \left[1 + \kappa |\mathbf{r}-\mathbf{r}'| \right ]^2.
\end{eqnarray}
Similarly, for electric dipoles of moment $\mathbf{p}$ and $\mathbf{p}'$ (in units of electronic charge) located at $\mathbf{r}$ and $\mathbf{r}'$, respectively, dipole-dipole interaction energy (in units of $k_B T$) is found to be 
\begin{eqnarray}
W_{dd}(\mathbf{r},\mathbf{p},\mathbf{r}',\mathbf{p}') &=& l_B\left[ A(|\mathbf{r} - \mathbf{r}'|)(\mathbf{p}.\mathbf{p}') - B(|\mathbf{r} - \mathbf{r}'|) \left[\mathbf{p}.(\mathbf{r} - \mathbf{r}')\right ]\left[\mathbf{p}'.(\mathbf{r} - \mathbf{r}')\right]  \right],
\end{eqnarray}
where 
\begin{eqnarray}
A(x)  &=& \frac{e^{-\kappa x}}{x^3} \left[1 + \kappa x \right ], \\
B(x) &=& \frac{e^{-\kappa x}}{x^5} \left[ 3 + 3\kappa x  + \kappa^2 x^2\right ].
\end{eqnarray}
For the freely rotating dipoles, the interaction energy becomes 
\begin{eqnarray}
W_{dd}^{f}(\mathbf{r},\mathbf{p},\mathbf{r}',\mathbf{p}') &=& -\frac{l_B^2}{3} p^2 p'^2 \frac{e^{-2\kappa |\mathbf{r} - \mathbf{r}'|}}{|\mathbf{r} - \mathbf{r}'|^6}C(\kappa |\mathbf{r} - \mathbf{r}'|),
\end{eqnarray}
where $C(x) = 1 + 2x + \frac{5}{3}x^2 + \frac{2}{3}x^3 + \frac{1}{6}x^4$ and 
$p' = |\mathbf{p}'|$. Note that putting $\kappa = 0$ in the above expressions, well-known functional form the bare charge-dipole and dipole-dipole interaction energies\cite{intermolecular_forces} are obtained. 

(iii) The radius of gyration of the polyzwitterionic chain is found to be dependent on an intricate interplay of charge-charge, charge-dipole and dipole-dipole interactions in addition to the short range excluded volume interactions. In this work, 
we show the possibility of an attraction dominated and a repulsion dominated regime. 
In the attraction dominated regime, dipolar interactions dominate over all the other interactions. For example, in the absence of the small ions, segment-segment interactions are described by the short range excluded volume interactions and the dipole-dipole interactions. In this case, the radius of gyration ($R_g^2 = Nll_1$) is to be obtained from the relation 
\begin{eqnarray}
       \frac{1}{l}  - \frac{1}{l_1} & = &  \frac{4}{3}\left(\frac{3}{2\pi}\right)^{3/2} \frac{w_{2}(Nl)^{1/2}}{l_1^{5/2}} + \frac{\nu_{eff}}{l_1^4}, \label{eq:size_salt_free_intro}
\end{eqnarray}
where $w_2$ and $\nu_{eff}$ are \textit{renormalized} two body and three body excluded volume parameters, given by Eqs. (~\ref{eq:w2}) and (~\ref{eq:nu}), respectively. These parameters depend on the Bjerrum's length and the magnitude of the dipole moment of the
zwitterionic monomers ($=p_m$). From Eq. (~\ref{eq:w2}), it is clear that for strong enough dipole-dipole interactions (characterized by the parameter $l_B^2 p_m^4$), $w_2$ may become negative even for good solvents (i.e., positive \textit{bare} excluded volume parameter $w$) and 
Eq. (~\ref{eq:size_salt_free_intro}) becomes the well-known relation describing a globule or a polymer chain in a poor solvent. In other words, our model predicts that a globule is the equilibrium state for strong enough dipolar interactions so that $w_2$ is negative. 

However, the presence of the small ions leads to the counterion adsorption on the zwitterionic monomers and the screening of the dipolar interactions. Taking into account the effects of the charge-charge and charge-dipole interactions in addition to the dipole-dipole interactions, the radius of gyration in the attraction dominated regime for the uniform distribution of charges on the zwitterionic sites is to be obtained from  
\begin{eqnarray}
       \frac{1}{l}  - \frac{1}{l_1} & = &  \frac{4}{3}\left(\frac{3}{2\pi}\right)^{3/2} \frac{w_{eff}(Nl)^{1/2}}{l_1^{5/2}} + \frac{\nu_{eff} - 12 w_{cd} \gamma l_1}{l_1^4} .\label{eq:size_salt_attraction_intro}
\end{eqnarray}
Here, $w_{cd}, \gamma, \nu_{eff}$ and $w_{eff}$ are given by Eqs. (~\ref{eq:wcd}), (~\ref{eq:gamma}), (~\ref{eq:nu}), and (~\ref{eq:w_eff}), respectively. From Eq. (~\ref{eq:size_salt_attraction_intro}) and (~\ref{eq:w_eff}), it is found that $l_1$ increases with the increase in $\kappa$ in the regime $l_1/l \ll 1$, which corresponds to a compact globule. Physically, the increase in $l_1$ with an increase in $\kappa$  corresponds to the opening up of the compact globule due to the screening of the attractive dipolar interactions. 
     
A key prediction of the model is the dependence of the degree of ionization of the zwitterionic monomers on the self-energy of the dipoles, charge-dipole and dipole-dipole pairs (given by Eqs. (~\ref{eq:ea_cc}), (~\ref{eq:ea_cd}) and (~\ref{eq:ea_dd}), respectively), formed as a result of the counterion adsorption. An asymmetric counterion adsorption on the positive and negative sites of the zwitterionic monomers is predicted for different adsorption energies of the two kinds of counterions. The adsorption energies (cf. Eq. (~\ref{eq:ea_cc})) are characterized by the local inhomogenieties in the dielectric constant near each kind of ions and the length of the dipole or ion-pair formed due to adsorption.  

For an asymmetric counterion adsorption, the repulsive charge-charge interactions may dominate over the attractive dipolar interactions. The cross-over from the attraction to repulsion dominated regime is characterized by $8 l_B(\alpha_+ - \alpha_-)^2 R_g^2/15 |w_{eff}|\simeq 1$. In the repulsion dominated regime, the radius of gyration is 
given by 
\begin{eqnarray}
       \frac{1}{l}  - \frac{1}{l_1} & = &  \frac{4}{3}\left(\frac{3}{2\pi}\right)^{3/2} \frac{w_{eff}(Nl)^{1/2}}{l_1^{5/2}} + \frac{4}{45}\left(\frac{6}{\pi}\right)^{1/2}\frac{l_B (\alpha_+ - \alpha_-)^2(Nl)^{3/2}}{l_1^{3/2}}\left(1 - \frac{\kappa^2 N l l_1}{7}\right)\label{eq:size_salt_repulsion_intro}
\end{eqnarray}
in the weak screening limit $\kappa R_g \rightarrow 0$ and 
\begin{eqnarray}
       \frac{1}{l}  - \frac{1}{l_1} & = &  \frac{4}{3}\left(\frac{3}{2\pi}\right)^{3/2} \left(w + \frac{4\pi l_B (\alpha_+ - \alpha_-)^2}{\kappa^2}\right)\frac{(Nl)^{1/2}}{l_1^{5/2}} \label{eq:size_salt_high2_intro}
\end{eqnarray}
in the strong screening limit  $\kappa R_g \rightarrow \infty$. Comparing Eqs. (~\ref{eq:size_salt_repulsion_intro}) and (~\ref{eq:size_salt_high2_intro}) with those for 
polyelectrolytes\cite{muthu87}, it is found that the polyzwitterionic chain behaves like a polyelectrolyte with the net charge equal to $e|\alpha_+ - \alpha_-|N$. Note that in this regime so that $l_1/l \gg 1$, the screening of electrostatic interactions lead to the decrease in the radius of gyration in contrast to the attraction dominated regime, where the screening leads to an increase in the radius of gyration. Also, note the dependence of the parameter characterizing the cross-over from an attraction to repulsion dominated regime (i.e., $l_B (\alpha_+ - \alpha_-)^2 R_g^2/|w_{eff}|$) on the salt concentration. Hence, the cross-over 
depends on the ionic strength related to $\kappa$ and the net charge on the polyzwitterionic chain.  
 
This paper is organized as follows: the formalism is presented in section ~\ref{sec:theory}, our results for the conformational characteristics of polyzwitterions are presented in
section ~\ref{sec:results}, and section ~\ref{sec:conclusions} contains our conclusions.

\section{Theory: Uniform Expansion Model }\label{sec:theory}
We consider a single flexible polyzwitterionic chain of
$N$ Kuhn segments, each with length $l$ in a spherical volume $\Omega = 4 \pi R^{3}/3$. The polyzwitterionic chain is represented
as a continuous curve of length $Nl$, and an arc length variable $s$
is used to represent any segment along the backbone so that $s \in [0,Nl]$ (Fig. ~\ref{fig:cartoon}). 
Also, we assume that
each segment has a dipole of length $r_d$ attached to it, which can rotate freely. Physically, this corresponds to a chain with zwitterionic side-groups attached to it, which carry both the monovalent positive and negative sites
separated by a spacer of length $r_d$. By taking the dipole length attached to each segment to be the same, we have
assumed that the distance between the positive and negative sites on the zwitterionic side groups is the same for each monomer and remains fixed irrespective of the conformational state of the chain. The distance is governed mainly by the chemistry of
the macromolecule and typically, it is around three to four methylene groups
(e.g., in poly-suphobetaines), which is about $0.5-0.7$ nm and amounts to a dipole moment of $24-34$ D for monovalent charged sites compared to a dipole moment of $1.85$ D for gaseous water (monomeric dipole-moment is represented by $p_m = er_d$ for univalently
charged groups). To have a general picture, we consider that there are $n_{c+}$ and $n_{c-}$
positive and negative counterions released by the negative and positive groups on the zwitterionic monomers, respectively. In addition to this, we assume that there are $n_{s+}$ and $n_{s-}$ positive and negative monovalent salt ions in the 
system. Overall, the system is electroneutral. We denote by $Z_j$ the valency (with sign) of the charged species $j$ and 
$j=+,-,c+,c-,s+,s-$ represent the positive, negative sites on the zwitterionic groups, positive, negative 
counterions from the zwitterionic monomers, positive and negative salt ions, respectively.

In order to study the effect of counterion adsorption on conformational characteristics 
in such a complicated multi-component system, we use the so-called ``two-state'' model\cite{muthu04}
for the counterions so that there are two populations of counterions. One population of the counterions
is free to enjoy the available volume (called the ``free'' counterions) and the other population is
``adsorbed'' on the backbone. However, the adsorbed counterions are allowed to move 
 along the backbone. In the literature, this kind of charge distribution has been referred to as
a ``permuted'' charge distribution\cite{borukhov98}. In the case of polyzwitterions, there are two 
kinds of counterions coming from the positive and negative charged sites of the 
zwitterionic monomers. We denote the degree of ionization of the 
negative and positive sites on the chain by $\alpha_+$ and $\alpha_-$, respectively, so that there 
are $-\alpha_+ NZ_{-}/Z_{c+}$ and $-\alpha_- NZ_{+}/Z_{c-}$ 
``free'' counterions coming from the negative and positive zwitterionic sites, respectively. In other words, 
there are $-(1-\alpha_{\pm})NZ_{\pm}/Z_{c\pm}$ ``adsorbed'' counterions on the chain and 
$e(\alpha_+ Z_{+} + \alpha_- Z_{-})N$ is the net charge on the chain, $e$ being the charge on an electron.

We compute the radius of gyration of a single polyzwitterionic chain in the 
presence of its counterions, the added salt ions, and the solvent (treated implicitly in this work 
as a uniform dielectric medium) using the variational formalism presented below. 

\subsection{Variational Formalism}\label{sec:variational}
Here, we present the variational formalism to compute the radius of gyration of a polyzwitterionic chain 
surrounded by small ions. In the presence of small ions, it can be shown that 
not only the charge-charge but also the charge-dipole and dipole-dipole interactions are screened (Appendix A). 
Using the functional forms for the screened interaction potentials derived in Appendix A, we can study the effective 
size of a polyzwitterionic chain in the presence of salt. However, the formation of small dipoles 
as a result of counterion adsorption complicates the situation. For the discussion here, we consider the case of monovalent salt and assume that 
the counterion adsorption of the positive and negative ions from the solution on the chain leads to dipoles having 
dipole moments $p_+$ and $p_-$, respectively.

There may also be bridging effects due to the presence of oppositely charged species along the backbone.
We can study this particular effect within uniform expansion model considered here 
by introducing another parameter $\alpha_b$, which is the fraction
of monomers involved in bridge formation. Bridging can be accounted for by adding a net attractive contribution 
to the excluded volume parameter\cite{arindam08}, which arises from a balance between ionic attractions 
and a conformational entropy penalty for bridges. However, in this work, we ignore the effect of bridging and 
focus on the isolated effect of dipolar interactions on the confomational characteristics of the zwitterionic chain.

Taking the dipole moment of each zwitterionic group
to be the same i.e., $p_m$ in magnitude, and modeling intra-group interactions by screened charge-charge, 
freely rotating charge-dipole and dipole-dipole interactions, the partition function can be written as
\begin{eqnarray}
       Z & = & \frac{Z_0 \exp[-E_a/k_BT]}{\mu} \int D[\mathbf{R}] \exp \left [- H_0\left\{\mathbf{R}\right\} - W\left\{\mathbf{R}\right\}\right ], \label{eq:partition_main}
\end{eqnarray}
where $H_0$ is the chain connectivity part given by
\begin{eqnarray}
 H_0\left\{\mathbf{R}\right\} &=& \frac {3}{2 l}\int_{0}^{Nl} ds\left(\frac{\partial \mathbf{R}(s)}{\partial s} \right )^{2}
\end{eqnarray}
and $W$ is the dimensionless interaction energy part (in units of $k_B T$, $k_B$ being Boltzmann's constant 
and $T$ being the temperature) written as
\begin{eqnarray}
W\left\{\mathbf{R}\right\} &=& \frac{w}{2l^2} \int_{0}^{Nl} ds \int_{0}^{Nl} ds' \delta\left[\mathbf{R}(s) - \mathbf{R}(s')\right] 
\nonumber \\
&& + \frac{\nu}{6l^3}  \int_{0}^{Nl} ds \int_{0}^{s} ds' \int_{0}^{s'} ds'' \delta\left[\mathbf{R}(s) - \mathbf{R}(s')\right]
\delta\left[\mathbf{R}(s') - \mathbf{R}(s'')\right] \nonumber \\
&+& \frac{1}{2l^2}\int_{0}^{Nl} ds \int_{0}^{Nl} ds' \left \{ w_{cc} V_{cc}\left [\mathbf{R}(s) - \mathbf{R}(s')\right] 
+ 2 w_{cd} V_{cd}\left[\mathbf{R}(s) - \mathbf{R}(s')\right ] \right . \nonumber \\
&& \quad \quad \left . + w_{dd} V_{dd}\left[\mathbf{R}(s) - \mathbf{R}(s')\right] \right \}.  \label{eq:inter_energy}
\end{eqnarray}

Here, $w$ is the conventional excluded volume parameter characterizing binary interactions, $\nu$ 
is the parameter characterizing ternary interactions, and $w_{cc}, w_{cd}$ and $w_{dd}$ are the prefactors 
determining the relative weightage of charge-charge, charge-dipole and dipole-dipole interactions, respectively (see Appendix A for the details). Explicitly, these are given by 
\begin{eqnarray}
w_{cc}  &=& l_B(\alpha_+ - \alpha_-)^2 \label{eq:wcc}\\
w_{cd} &=& -\frac{l_B^2}{6} (\alpha_+ + \alpha_- - 2\alpha_+\alpha_-)\left[\alpha_+\alpha_-\left(\frac{p_m}{e}\right)^2 + (1-\alpha_+)\left(\frac{p_+}{e}\right)^2 + (1-\alpha_-)\left(\frac{p_-}{e}\right)^2\right], \nonumber \\
&& \label{eq:wcd} \\
w_{dd} &=& -\frac{l_B^2}{6} \left[\alpha_+\alpha_-\left(\frac{p_m}{e}\right)^2 + (1-\alpha_+)\left(\frac{p_+}{e}\right)^2 + (1-\alpha_-)\left(\frac{p_-}{e}\right)^2\right]^2, \label{eq:wdd}
\end{eqnarray}
where $l_B = e^{2}/4\pi \epsilon_0 \epsilon_r k_B T$ is Bjerrum's length written in terms of the 
relative permittivity $\epsilon_r$ of the medium, $\epsilon_0$ being the permittivity of vacumm.

Also, $V_{cc}, V_{cd}$ and $V_{dd}$ are the screened charge-charge, freely rotating charge-dipole and dipole-dipole 
interaction potentials, respectively (see Appendix A), given by
\begin{eqnarray}
V_{cc}(\mathbf{x}) &=&  \frac{\exp \left [ - \kappa x\right ]}{x},\label{eq:cc_potential} \\
V_{cd}(\mathbf{x}) &=& \frac{\exp \left [ - 2\kappa x\right ]}{x^4} \left(1 + \kappa x\right)^2, \label{eq:cd_potential} \\
V_{dd}(\mathbf{x}) &=&  \frac{\exp \left [ - 2\kappa x\right ]}{x^6} C\left(\kappa x\right), \label{eq:dd_potential}
\end{eqnarray}
where $x = |\mathbf{x}|$ and $C(x)$ is defined in Eq. (~\ref{eq:c_equation}) by 
\begin{eqnarray}
C(x) &=& 1 + 2x + \frac{5}{3}x^2 + \frac{2}{3}x^3 + \frac{1}{6}x^4. \label{eq:c_equation_main}
\end{eqnarray}
Note that the prefactors containing the magnitude of the dipole moments and the charges in the expressions for the charge-charge, charge-dipole and dipole-dipole interaction potentials are taken away in the definition of $w_{cc}, w_{cd}$ and $w_{dd}$, respectively,
 for the writing purposes. Furthermore, $\kappa^2 = 4\pi l_B \sum_j Z_j^2 n_j/\Omega$, where $j=c+,c-,s+,s-$, so that $1/\kappa$ is the Debye's screening length, and 
$n_{c+} = \alpha_+ N$ and  $n_{c-} = \alpha_- N$ are the number of positive and negative counterions, respectively.

Due to the attractive nature of charge-dipole and dipole-dipole interactions, there might be a chain collapse. 
To stabilize against a collapsed conformational 
state of the chain, repulsive ternary interactions\cite{muthu84,birshtein,dua05} characterized by $\nu > 0$
are also taken into account into Eq. ~\ref{eq:inter_energy}. 
Also, the effect of solvent is modeled by ignoring the interactions between the charged species and
solvent molecules and carrying out weak inhomogeneity expansion for the solvent density  
(also known as random-phase approximation (RPA)), which leads to a renormalization of the excluded volume parameter by 
$w = (1/(1-\phi_p) - 2\chi_{ps})l^3$, $\phi_p = Nl^3/\Omega$ being the volume fraction of the monomers.

In Eq. (~\ref{eq:partition_main}), $Z_0$ is the partition function for small ions at the level of 
one-loop equivalent to Debye-H\"{u}ckel or RPA, which captures the 
effect of fluctuations in the number density of small ions. Explicitly, it is given by 
\begin{eqnarray}
-\ln Z_{0} &=& \sum_{j=c+,c-,s+,s-} n_j (\ln \frac{n_j}{\Omega} - 1) - \frac{\Omega \kappa^3}{12 \pi}, \label{eq:si}
\end{eqnarray}
where the first term takes into account the translational entropy of the small ``free'' ions  and the second term 
is responsible for the fluctuations in the density of the small ions. Also, $E_a$ and $\mu$ are the parts of the so called chemical free energy\cite{overbeek90} of the system, originating from the 
``adsorbed'' counterions. $E_a$ is the energetic part of the chemical free energy and  
includes the self-energy of the ion-pairs ($E_a^{cc}$), charge-dipole ($E_a^{cd}$) and 
dipole-dipole ($E_a^{dd}$) pairs, given by
\begin{eqnarray}
\frac{E_{a}}{k_B T} &=& \frac{E_{a}^{cc}}{k_B T} + \frac{E_{a}^{cd}}{k_B T} + \frac{E_{a}^{dd}}{k_B T} \label{eq:ea}\\
\frac{E_{a}^{cc}}{k_B T} &=& - \left[(1-\alpha_+)\delta_+  + (1-\alpha_-)\delta_-\right] Nl_{B}/l, \label{eq:ea_cc}\\
\frac{E_{a}^{cd}}{k_B T} &=& - \left [(1-\alpha_+)\alpha_- \left(\frac{p_+}{e}\right)^2  + (1-\alpha_-)\alpha_+ \left(\frac{p_-}{e}\right)^2\right] \frac{N l_{B}^2}{6 r_d^4}, \label{eq:ea_cd}\\
\frac{E_{a}^{dd}}{k_B T} &=& - (1-\alpha_+)(1-\alpha_-)\left(\frac{p_+}{e}\right)^2\left(\frac{p_-}{e}\right)^2\frac{Nl_{B}^2}{3 r_d^6}. \label{eq:ea_dd}
\end{eqnarray}

In these expressions for the self-energies of different kinds of pairs, the parameters 
$\delta_{\pm} = \epsilon l/\epsilon_{l\pm}d_{\pm}$, capture the effect of the deviation of the local dielectric constant near the polyzwitterioinic chain ($\epsilon_{l\pm}$) from the bulk value ($\epsilon$). Also, $d_\pm$ represents the length of the dipole formed due to ion-pairing by positive or negative counterion (i.e., $p_\pm = ed_\pm$).

$\mu$ is the number of ways of distributing the ``adsorbed'' counterions (say $N_+, N_-$) among $N$ charged sites,
given by $\mu =  \frac{N!}{N_+!(N-N_+)!} \frac{N!}{N_-!(N-N_-)!}$. The explicit expression for $\mu$ can be 
used to compute the entropic part ($S_a$) of the chemical free energy by  
\begin{eqnarray}
\frac{-TS_{a}}{k_B T} &=& -\ln \mu  = N\left[ \alpha_+\log \alpha_+ + (1-\alpha_+)\log (1-\alpha_+) \right . \nonumber \\
&& \left. + \alpha_-\log \alpha_- + (1-\alpha_-)\log (1-\alpha_-)\right],\label{eq:sa}
\end{eqnarray}
where $\alpha_{\pm} = N_{\pm}/N$ and Stirling's approximation $\ln n! \simeq n\ln n - n$ has been used 
in writing Eq. (~\ref{eq:sa}). We must stress here that these contributions to the chemical 
free energy of the system are independent of the conformation of the chain. However, the 
size of the chain depends on the chemical free energy in an implicit way through $\alpha_+$ and 
$\alpha_-$. 

Using the variational method presented in Ref. \cite{muthu87} along with the partition function given in 
Eq. (~\ref{eq:partition_main}), the conformational characteristics of the polyzwitterionic chain can be 
computed by approximating it by an \textit{effective} Gaussian chain with Kuhn step length $l_1$,
where $l_1$ depends on the various intra-chain interactions. Also, the radius of gyration ($R_g$) 
of the chain is related to the effective step length $l_1$ by the relation $R_g^2 = Nll_1/6$. 
Following Ref. $26$, 
the effective step length $l_1$ is given by the relation
  
\begin{eqnarray}
       \frac{1}{l}  - \frac{1}{l_1} & = &  wI_{ww} + w_{cc}I_{cc} + w_{cd}I_{cd} + w_{dd}I_{dd} + \frac{\eta \nu}{l_1^4}, \label{eq:chain_size}
\end{eqnarray}
where $\eta$ is given by 
\begin{eqnarray}
\eta &=& \left(\frac{3}{2\pi}\right)^3\int_{0}^{1} ds \int_{0}^{s} ds' \int_{0}^{s'} ds''\frac{(s-s'')}{\left[(s-s')(s'-s'')\right]^{3/2}}, \label{eq:eta}
\end{eqnarray}
which is divergent and needs to be regularized (see Refs. $31-33$ 
for the derivation of this term ). For the discussion in this paper, we 
have taken $\eta$ to be a positive constant obtained after carrying out the regularization. Furthermore, 
\begin{eqnarray}
I_{ww} &=& \frac{1}{18 N l}\int_{0}^{Nl} ds \int_{0}^{Nl} ds' (s-s')^2 \int \frac{d^3 k}{(2\pi)^3} k^2 \exp \left[ -k^2 l_1 |s-s'|/6\right ], \label{eq:iww}\\
I_{cc} &=& \frac{1}{18 N l}\int_{0}^{Nl} ds \int_{0}^{Nl} ds' (s-s')^2 \int \frac{d^3 k}{(2\pi)^3} V_{cc}(k) k^2 \exp \left[ -k^2 l_1 |s-s'|/6\right ],\label{eq:icc}\\
I_{cd} &=& \frac{1}{18 N l}\int_{0}^{Nl} ds \int_{0}^{Nl} ds' (s-s')^2 \int \frac{d^3 k}{(2\pi)^3} 2 V_{cd}(k) k^2 \exp \left[ -k^2 l_1 |s-s'|/6\right ],\label{eq:icd}\\
I_{dd} &=& \frac{1}{18 N l}\int_{0}^{Nl} ds \int_{0}^{Nl} ds' (s-s')^2 \int \frac{d^3 k}{(2\pi)^3} V_{dd}(k)  k^2 \exp \left[ -k^2 l_1 |s-s'|/6\right ]\label{eq:idd}.
\end{eqnarray}

In above equations, $V_{cc}(k), V_{cd}(k)$ and $V_{dd}(k)$ are the Fourier components of the screened charge-charge, 
charge-dipole and dipole-dipole interaction potentials given in 
Eqs. (~\ref{eq:cc_potential} - ~\ref{eq:dd_potential}), respectively.
Introducing a short distance cut-off in the real space ($ = \lambda \rightarrow 0$, so that $\lambda$ has the units of 
length) to regularize the divergent integrals in 
the computations of Fourier transforms for the screened charge-dipole and dipole-dipole interaction 
potentials, explicit expressions for $V_{cc}(k), V_{cd}(k)$ and $V_{dd}(k)$ are given by
\begin{eqnarray}
V_{cc}(k) &=& \frac{4\pi}{k^2 + \kappa^2}, \\
V_{cd}(k) &=& 4\pi \left[\frac{3}{4\lambda}  + \kappa - \frac{k^2 + 2\kappa^2}{2k}\arctan\left(\frac{k}{2\kappa}\right)\right],\\
V_{dd}(k) &=& 4\pi \left[\frac{7}{48 \lambda^3} + \frac{\kappa}{3\lambda^2} + \frac{\kappa^2}{6\lambda} - \frac{k^2}{16 \lambda} - \frac{\kappa k^2}{12} \right . \nonumber \\
&& \left .  + \frac{1}{24 k}\left (k^4 + 16\kappa^2 k^2 + 4\kappa^4\right )\arctan\left(\frac{k}{2\kappa}\right)\right].
\end{eqnarray}

Evaluation of $I_{ww}$ and $I_{cc}$ has already been carried out in the literature\cite{edwards_singh,muthu87}. 
In particular,
\begin{eqnarray}
I_{ww} &=& \frac{4}{3}\left(\frac{3}{2\pi}\right)^{3/2} \frac{(Nl)^{1/2}}{l_1^{5/2}}, \label{eq:iww_final}\\
I_{cc} &=& \frac{4}{45}\left(\frac{6}{\pi}\right)^{1/2}\frac{(Nl)^{3/2}}{l_1^{3/2}}\Theta_0^p (a), \label{eq:icc_final}
\end{eqnarray}
where
  \begin{eqnarray}
\Theta_0^p (a) &=&  \frac{15 \sqrt{\pi}}{2 a^{5/2}}\left(a^2 - 4a + 6\right)\exp(a) \mbox{erfc} (\sqrt{a})
+ \frac{15}{\sqrt{\pi}}\left(  - \frac{3\pi}{a^{5/2}} - \frac{\pi}{a^{3/2}}  + \frac{6\sqrt{\pi}}{a^2}\right),
\end{eqnarray}
and  $a = \kappa^2 R_g^2 = \kappa^2 N l l_1/6$.

$I_{cd}$ and $I_{dd}$ can be evaluated using the expressions for $V_{cd}(k)$ and $V_{dd}(k)$, respectively.
Unfortunately, analytical evaluations of these integrals are not possible for arbitrary values of $\kappa$. 
However, in the limiting cases of $b = 4a \rightarrow 0$ and $b\rightarrow \infty$, the integrals can be carried out 
analytically (see Appendix B for details). Physically, the limits of $b\rightarrow 0$ and $b\rightarrow \infty$ correspond to the collapsed globule state in low salt concentrations and expanded coil for moderate salt concentrations, respectively. 

For these limiting cases, $I_{cd}$ and $I_{dd}$ are given by
\begin{equation}
 I_{cd}  = \left \{ \begin{array}{ll}
\frac{4}{3}\left(\frac{3}{2\pi}\right)^{3/2} \frac{(Nl)^{1/2}}{l_1^{5/2}}\frac{3\pi}{\lambda}
- \frac{12 \gamma}{l_1^3} + \frac{24\sqrt{b}}{\sqrt{\pi} l_1^3}, & b \rightarrow 0 \\
  &  \\
 \frac{4}{3}\left(\frac{3}{2\pi}\right)^{3/2} \frac{(Nl)^{1/2}}{l_1^{5/2}}
\frac{3\pi}{\lambda}  - \frac{12}{l_1^3} + \frac{6\sqrt{b}}{\sqrt{\pi} l_1^3}, & b \rightarrow \infty
\end{array}
   \right . \label{eq:icd_lim}
\end{equation}
and 
\begin{equation}
I_{dd}  = \left \{ \begin{array}{ll}
\frac{4}{3}\left(\frac{3}{2\pi}\right)^{3/2} \frac{(Nl)^{1/2}}{l_1^{5/2}} \left[4\pi \left(\frac{7}{48 \lambda^3} + \frac{\kappa}{3\lambda^2} + \frac{\kappa^2}{6\lambda} \right )\right ] + 30\sqrt{\frac{6}{\pi}}\frac{\kappa}{(Nl)^{1/2}l_1^{7/2}}  & \\
+ \frac{18 \xi}{N l l_1^{4}} + \frac{16 \kappa^2 \gamma}{l_1^3}, & b \rightarrow 0 \\
  &  \\
 \frac{4}{3}\left(\frac{3}{2\pi}\right)^{3/2} \frac{(Nl)^{1/2}}{l_1^{5/2}} \left[4\pi \left(\frac{7}{48 \lambda^3} + \frac{\kappa}{3\lambda^2} + \frac{\kappa^2}{6\lambda} \right )\right ]
- \frac{175}{4}\sqrt{\frac{6}{\pi}}\frac{\kappa}{(Nl)^{1/2}l_1^{7/2}}  & \\
+ \frac{\kappa^2}{l_1^3} \left ( 15  +  \frac{\sqrt{b}}{\sqrt{\pi}}\right ), & b \rightarrow \infty
\end{array}
   \right .  \label{eq:idd_lim}
\end{equation}
respectively. In these equations, $\gamma$ and $\xi$ are given by 
\begin{eqnarray}
       \gamma & = &  \int_{0}^{1}ds \int_{0}^{s}ds' \frac{1}{s-s'} \label{eq:gamma}
\end{eqnarray}
and 
\begin{eqnarray}
       \xi & = &  \int_{0}^{1}ds \int_{0}^{s}ds' \frac{1}{\left(s-s'\right)^2}, \label{eq:xi}
\end{eqnarray}
respectively. Note that like $\eta$ in Eq. (~\ref{eq:eta}), $\gamma$ and $\xi$ are also divergent and need to be 
regularized. Also, the origin of the divergences in the expressions for $\eta, \gamma$ and $\xi$ lie in the use of continuous integrals while writing the interaction energy 
in Eq. (~\ref{eq:inter_energy}). These kinds of divergences often appear in the coarse-grain models and can 
be regularized by introdcing an appropriate cut-off. However, in this paper, we do not explicitly regularize these 
quantities, but treat them as constants obtained after appropriate regularization.       

Using Eqs. (~\ref{eq:chain_size}), (~\ref{eq:iww_final}),  (~\ref{eq:icc_final}),  (~\ref{eq:icd_lim}), and (~\ref{eq:idd_lim}), the 
effect of dipolar interactions and the added salt 
on the radius of gyration of the polyzwitterionic chain can be studied in the limiting cases. 
Note that for the limiting cases considered here 
\begin{equation}
\Theta_0^p (a) = \left \{ \begin{array}{ll}
1 - 6a/7, & a \rightarrow 0 \\
 15/2a, & a \rightarrow \infty.
\end{array}
   \right .  \label{eq:theta_cc_lim}
\end{equation}
In the next section, we present the conformational characteristics of a single polyzwitterionic chain 
in a salt-free as well as salty environment using the theoretical formalism presented above.
  

\section{Conformational Characteristics}\label{sec:results}
\subsection{Salt-Free Dipolar Polyzwitterionic Chain}
Consider the case of a single purely dipolar polyzwitterionic chain in a salt-free environment, 
so that there are no added salt ions as well as counterions from the chain (i.e., $\alpha_+ = \alpha_- = 1$). 
In this scenario, the monomers of the polyzwitterionic chain interact with each other by bare dipole-dipole interactions 
along with the excluded volume interactions. 
Putting $\alpha_+ = \alpha_- = 1$ in Eqs. (~\ref{eq:wcc} - ~\ref{eq:wdd}) and $\kappa = 0$ in 
Eq. (~\ref{eq:idd_lim}),  Eq. (~\ref{eq:chain_size}) becomes
\begin{eqnarray}
       \frac{1}{l}  - \frac{1}{l_1} & = &  \frac{4}{3}\left(\frac{3}{2\pi}\right)^{3/2} \frac{w_{2}(Nl)^{1/2}}{l_1^{5/2}} + \frac{\nu_{eff}}{l_1^4}, \label{eq:size_salt_free}
\end{eqnarray}
where $w_2$ and $\nu_{eff}$ are \textit{renormalized} two body and three body excluded volume parameters, given by 
\begin{eqnarray}
       w_{2} & = &  w - \frac{l_B^2}{6}\left(\frac{p_m}{e}\right)^4 \left ( \frac{7 \pi}{12 \lambda^3}\right ), \label{eq:w2}
\end{eqnarray}
and
\begin{eqnarray}
       \nu_{eff} & = &  \eta \nu - \frac{3 l_B^2}{N l}\left(\frac{p_m}{e}\right)^4 \xi , \label{eq:nu}
\end{eqnarray}
respectively. 

In the expression for the \textit{renormalized} excluded volume parameter $w_2$, the first term is the 
excluded volume parameter renormalized due to the presence of solvent, i.e., $w =  (1/(1-\phi_p)-2\chi_{ps})l^3$,
and the negative second term represents the effect of dipole-dipole interactions on the short range interaction strength. 
This implies that the dipolar interactions add an attractive component to the short-range
excluded volume interactions, which may be purely repulsive or attractive depending on the solvent quality.

Conformational characteristics and in turn, the solubility of
the purely dipolar polyzwitterion molecules in a salt-free environment can be understood by estimating the
sign of the the renormalized excluded volume parameter $w_2$.
Using well-known concepts\cite{scalingbook}, if $w_2 > 0$, the polyzwitterionic chain behaves like a
polymer in a good solvent and its radius of gyration is greater than that of the corresponding phantom chain (whose connected segments don't interact with 
each other). However, if $w_2<0$, the chain behaves like it is in poor solvent conditions and its radius 
of gyration is smaller than that of the phantom chain.
Similarly, if $w_{2}=0$, then an equivalent of theta-solvent condition for neutral polymers
can be envisaged and the radius of gyration of the chain equals that of the phantom chain. From Eq. (~\ref{eq:w2}), it is clear that the dipole-dipole interactions tend to reduce the
solvent quality and leads to a shrinkage of the chain.

Similarly, the three body interaction term gets renormalized due to the presence of dipolar 
interactions(cf. Eq. (~\ref{eq:nu})).
However, the reduction in the ternary interaction term due to the dipole-dipole interactions is
small, of the order $\sim 1/N$. So, for infinitely long chains, the renormalization of the third body
interaction term is negligible. Due to the repulsive nature of the ternary interaction term, a reduction 
in this term because of the dipolar interactions aids in the shrinkage of the chain.  

The effect of the dipole-dipole interactions on the binary and ternary interaction terms can be 
used to obtain some scaling laws for the radius of gyration of a polyzwitterionic chain. It has been already shown\cite{scalingbook,muthu84,birshtein,sanchez79,globule_review} that in the presence of attractive two body interactions, 
the chain tends to shrink. However, the shrinkage is unfavorable because of the loss in conformational entropy 
as well as the repulsive ternary interactions\cite{muthu84,birshtein}. In the collapsed globule regime, ternary repulsive interactions dominate over the conformational entropy. On the other hand, in the expanded coil regime, 
the chain conformational entropy dominates over the repulsive ternary interactions.  
The cross-over from one regime to the other can be estimated by balancing the chain conformational entropy 
with the ternary repulsive term in Eq. (~\ref{eq:size_salt_free}). 

In the expanded coil regime so that $w_2 > 0$, the equilibrium radius of gyration can be estimated by balancing the left hand 
side with the first term on the right hand side in Eq. (~\ref{eq:size_salt_free}), 
which gives Flory's result\cite{scalingbook,flory_book} for $R_g = \sqrt{N l l_1/6} \sim (w_2)^{1/5}N^{3/5}$. From the dependence of $w_2$ 
on $p_m$ and $l_B$ in Eq. (~\ref{eq:w2}), it is clear that the radius of gyration of the chain 
decreases with an increase in the monomeric dipole moment ($p_m$) or the Bjerrum length. 

In the poor solvent regime\cite{scalingbook,muthu84,birshtein,sanchez79,globule_review} for the polyzwitterionic chain so that $w_2$ is negative, the equilibrium radius of gyration can be estimated by balancing the first and second term on the right 
hand side in Eq. (~\ref{eq:size_salt_free}), which gives $R_g = \sqrt{N l l_1/6} \sim (\nu_{eff}/|w_2|)^{1/3}N^{1/3}$. 
Due to an increase in $|w_2|$ and a decrease in $\nu_{eff}$ on increasing the strength of dipole-dipole interactions, 
the collapse of the dipolar polyzwitterionic chain is stronger as the monomeric dipole moment ($p_m$) or 
the Bjerrum length is increased. 

Hence, dipole-dipole interactions always cause a shrinkage in the radius of the chain and sometimes may lead 
to a chain collapse. In other words, the dipole-dipole interactions reduce the solvent quality for the polyzwitterionic chain, which is consistent with the experimentally observed insolubility of polyzwitterionic molecules in aqueous solutions.     

\subsection{Polyzwitterionic Chain in the Presence of Added Salt}
Interpretation of the quantitative relations between the radius of gyration of a polyzwitterionic chain 
in the presence of its own counterions or added salt is facilitated by noticing that in general, small ions have 
two effects on the conformational characteristics of the chain. First is the screening of the charge-charge, 
charge-dipole and dipole-dipole interactions presented in Eqs. (~\ref{eq:cc_potential}-~\ref{eq:dd_potential}).
Second is the possibility of counterion adsorption on the zwitterionic side-groups, which dictates 
considering the relative importance of the three different kinds of interactions (cf. Eqs. (~\ref{eq:wcc}-~\ref{eq:wdd})). 

The role of counterion adsorption on the polyzwitterionic chain demands systematic attention 
using the theoretical tools presented here. However, we can infer some important results 
from well-known counterion adsorption phenomena for a flexible 
polyelectrolyte chain. For a single flexible polyelectrolyte chain, it has been shown\cite{muthu04,huber_review,
rajeev_arindam} that
there is a non-trivial dependence of the degree of ionization on the added salt concentration and the 
parameter characterizing the binding energy of the counterions on to the chain (i.e., an analogue 
of $\delta_{\pm}$  for polyelectrolytes). Furthermore, the degree of ionization at equilibrium has been shown to be 
determined mainly as an interplay of counterion adsorption energy ($E_a$), translational entropy and correlation 
energy of the ``free'' ions ($-\ln Z_0$) and the translational entropy of the ``adsorbed'' counterions ($S_a$). 
In fact, the degree of ionization decreases with increasing salt concentration and counterion binding energy. 

The counterion adsorption phenomenon is richer in the case of polyzwitterions due to the presence of two kinds of 
charged sites on the chain. In the case of polyelectrolytes, it has been shown that the chain conformational entropy 
has almost no effect on the equilibrium degree of ionization in the good solvent regime. Assuming the 
chain conformational entropy 
to have the similar effect in the case of polyzwitterions also, it is clear from 
Eqs. (~\ref{eq:si}), (~\ref{eq:ea}) and (~\ref{eq:sa}) that the equilibrium degree of ionization depends 
on the salt concentration, the dipole length $r_d$, and the 
parameters $\delta_{\pm}$. This means the asymmetry in the degree of counterion adsorption 
on the polyzwitterionic sites depends on the parameters $\delta_{\pm}$. To demonstrate this point, we have 
minimized the free energy with respect to $\alpha_{\pm}$, retaining only $E_a, -\ln Z_0$ and $-TS_a$. The results 
for a particular set of parameters relevant to polyzwitterions are shown in Fig. ~\ref{fig:asymmetry}. For equal 
values of $\delta_+$ and $\delta_-$, both positive and negative counterions adsorb 
equally on the corresponding zwitterionic sites so that there is no asymmetry in the 
degrees of counterion adsorption (Fig. ~\ref{fig:asymmetry}(a) ). On the other hand, increasing the parameter $\delta_{\pm}$ causes more counterions to adsorb on 
the polyzwitterionic sites and hence, enhances the counterion adsorption asymmetry. This particular point 
is demonstrated by Figs. ~\ref{fig:asymmetry}(b) and ~\ref{fig:asymmetry}(c), where the larger value of 
$\delta_{\pm}$ leads to an enhanced adsorption of the corresponding counterions compared to the others. 
The decrease in $\alpha_{\pm}$ 
with the increase in Bjerrum length in these figures is consistent with the known result from polyelectrolytes.  

Furthermore, the parameters $\delta_{\pm}$ characterize the binding energies of positive and negative counterions 
on the polyzwitterionic chain. Also, note that the parameters $r_d$ and $\delta_{\pm}$ 
embody a chemical specificity; dependence of the degree of ionization on these parameters means that local size and chemical details of the zwitterionic groups and the counterions play an important 
role in the counterion adsorption phenomenon. These concepts of counterion adsorption will be seen to have implications for the conformational characteristics of polyzwitterionic chains as described below. 

\subsubsection{Weak screening limit i.e., $b\rightarrow 0$}
In the presence of ``free'' ions near the polyzwitterionic chain so that $b\rightarrow 0$, Eq. (~\ref{eq:chain_size}) becomes
\begin{eqnarray}
       \frac{1}{l}  - \frac{1}{l_1} & = &  \frac{4}{3}\left(\frac{3}{2\pi}\right)^{3/2} \frac{w_{eff}(Nl)^{1/2}}{l_1^{5/2}} + \frac{4}{45}\left(\frac{6}{\pi}\right)^{1/2}\frac{w_{cc}(Nl)^{3/2}}{l_1^{3/2}}\left(1 - \frac{\kappa^2 N l l_1}{7}\right) \nonumber \\
       && + \frac{(- 12 w_{cd} + 16 w_{dd}\kappa^2)\gamma}{l_1^3}  + \frac{\nu_{eff}}{l_1^4} + 30\sqrt{\frac{6}{\pi}}\frac{w_{dd}\kappa}{(Nl)^{1/2}l_1^{7/2}}, 
\label{eq:size_salt}
\end{eqnarray}
where $\gamma$ is a constant given by Eq. ~\ref{eq:gamma} and $w_{eff}$ is the \textit{effective} excluded volume parameter in the presence of small ions ( which may be the counterions from the chain or added salt ions or both), given by

\begin{eqnarray}
       w_{eff} & = &  w + w_{cd}\left[\frac{3\pi}{\lambda} + 8\pi \kappa \right] + w_{dd}\left[4\pi \left( \frac{7}{48 \lambda^3} + \frac{\kappa}{3\lambda^2} + \frac{\kappa^2}{6\lambda} \right )\right ]. \label{eq:w_eff}
\end{eqnarray}

In the above expression for $w_{eff}$, the first term is the excluded volume 
parameter \textit{renormalized} due to the presence of solvent. The second term represents the
the effect of charge-dipole interactions on the short ranged excluded volume interactions and is negative ($w_{cd}<0$). 
The $\kappa$ dependent part in this term comes from the screening of charge-dipole interactions 
by the small ions. The third term is 
the already mentioned (see Eq. (~\ref{eq:w2})) dipole-dipole interaction term and has additional contributions 
in this context due to ion screening. Note that the decrease in $w_{eff}$ due to the screening of 
charge-dipole and dipole-dipole interactions does not imply that the addition of the salt leads to the shrinkage 
of the chain.  

The presence of ``free'' ions near a polyzwitterionic chain, the chain conformational characteristics and in turn, the 
solubility of polyzwitterions is determined by an intricate interplay of attractive dipolar 
(charge-dipole and dipole-dipole) and repulsive 
charge-charge interactions along with the \textit{bare} binary and ternary interactions. 
If the attractive charge-dipolar interactions dominate over the repulsive charge-charge interactions 
(i.e., $8w_{cc}R_g^2/15 |w_{eff}| \ll 1$), 
then Eq. (~\ref{eq:size_salt}) becomes  

\begin{eqnarray}
       \frac{1}{l}  - \frac{1}{l_1} & = &  \frac{4}{3}\left(\frac{3}{2\pi}\right)^{3/2} \frac{w_{eff}(Nl)^{1/2}}{l_1^{5/2}} + \frac{\nu_{eff} - 12 w_{cd} \gamma l_1}{l_1^4} ,\label{eq:size_salt_attraction}
\end{eqnarray}
which is written after neglecting the charge-charge and the dipole-dipole interaction 
terms on the right hand side in Eq. (~\ref{eq:size_salt}). Physically, this equation corresponds to a compact globule state for the polyzwitterionic chain in a regime, where charge-dipole interactions dominate over the repulsive charge-charge interactions.

The effect of screening of the dipolar 
interactions appear in the form of negative contributions containing $\kappa$ on the right hand side 
in Eq. (~\ref{eq:size_salt_attraction}) (in the expression for $w_{eff}$) and an expansion of the chain with 
the increase in $\kappa$ can be inferred 
by noting that in the limit of $l_1/l \ll 1$, the left hand side in Eq. (~\ref{eq:size_salt_attraction}) is 
negative and an increase in $l_1/l$ makes it more negative. 
However, it must be kept in mind that the screening of the attractive dipolar interactions alone 
can not lead to an expanded coil state past the theta point. In other words, the screening of the dipolar interactions 
can only lead to an opening up of the compact globular state.   

On the other hand, one can imagine that an addition of salt may lead to sufficient asymmetric counterion adsorption
($|\alpha_+ -  \alpha_-|$) on the polyzwitterionic chain so that the repulsive interactions become comparable to the attractive dipolar interactions (i.e., $8w_{cc}R_g^2/15 |w_{eff}| \simeq 1$). In this regime, the polyzwitterionic chain takes an expanded coil conformation (note that this may happen in the low salt concentrations so that $b\rightarrow 0$) and Eq. (~\ref{eq:size_salt}) becomes

\begin{eqnarray}
       \frac{1}{l}  - \frac{1}{l_1} & = &  \frac{4}{3}\left(\frac{3}{2\pi}\right)^{3/2} \frac{w_{eff}(Nl)^{1/2}}{l_1^{5/2}} + \frac{4}{45}\left(\frac{6}{\pi}\right)^{1/2}\frac{w_{cc}(Nl)^{3/2}}{l_1^{3/2}}\left(1 - \frac{\kappa^2 N l l_1}{7}\right),\label{eq:size_salt_repulsion}
\end{eqnarray}
which follows by neglecting the last three terms on the right hand side in Eq. (~\ref{eq:size_salt}).
 
Comparing Eq. (~\ref{eq:size_salt_repulsion}) with that for a flexible polyelectrolyte\cite{muthu87}, it is 
clear that in this regime, the polyzwitterionic chain behaves like a polyelectrolyte 
chain with charge $e|\alpha_+ -  \alpha_-|N$. An important point to note is the dependence of the counterion adsorption 
asymmetry on the specificity of the zwitterionic groups, the counterions and the added salt. Furthermore, the 
theoretical prediction of the cross-over from the 
attraction dominated regime to the polyelectrolyte regime requires a minimum 
amount of salt so that $8w_{cc}R_g^2/15 |w_{eff}| \simeq 1$. This is qualitatively in agreement with the experiments\cite{kudaibergenov_review06,salamone78}, where 
a minimum amount of salt is observed to achieve solubility of the polyzwitterions. 

Hence, the addition of salt leads to an opening up of the collapsed globule in the 
attraction dominated regime and may cause an expansion due to the counterion adsorption asymmetry on the 
polyzwitterionic chain. The expansion of the polyzwitterionic chain due to the counterion adsorption asymmetry 
may also lead to the other limiting case, where $b\rightarrow \infty$, which is 
presented below. 

\subsubsection{Strong screening limit i.e., $b\rightarrow \infty$}
In the strong screening limit so that $b\rightarrow \infty$, Eq. (~\ref{eq:chain_size}) becomes
\begin{eqnarray}
       \frac{1}{l}  - \frac{1}{l_1} & = &  \frac{4}{3}\left(\frac{3}{2\pi}\right)^{3/2} \frac{w_{eff}(Nl)^{1/2}}{l_1^{5/2}} 
       + \frac{(- 12 w_{cd} + 15 w_{dd}\kappa^2)}{l_1^3} 
+ \frac{\eta \nu}{l_1^4} - \frac{175}{4}\sqrt{\frac{6}{\pi}}\frac{w_{dd}\kappa}{(Nl)^{1/2}l_1^{7/2}},\nonumber \\
&& \label{eq:size_salt_high}
\end{eqnarray}
where $w_{eff}$ is the \textit{effective} excluded volume parameter in this limit, given by 
\begin{eqnarray}
       w_{eff} & = &  w + \frac{4\pi w_{cc}}{\kappa^2} + w_{cd}\left[\left(\frac{3\pi}{\lambda}\right) + 2\pi \kappa \right] + w_{dd}\left[4\pi \left(\frac{7}{48 \lambda^3} + \frac{\kappa}{3\lambda^2} + \frac{\kappa^2}{6\lambda} \right )  + \frac{\pi \kappa^3}{3} \right ]. \nonumber \\
&& \label{eq:w_eff_high}
\end{eqnarray}

Noting that this limit is attained due to the dominance of the charge-charge interactions over the charge-dipole and dipole-dipole interactions, Eq. (~\ref{eq:size_salt_high}) can be rewritten after neglecting the terms containing $w_{cd}$ and $w_{dd}$ so that Eqs. (~\ref{eq:size_salt_high}) and (~\ref{eq:w_eff_high}) become 

\begin{eqnarray}
       \frac{1}{l}  - \frac{1}{l_1} & = &  \frac{4}{3}\left(\frac{3}{2\pi}\right)^{3/2} \frac{w_{eff}(Nl)^{1/2}}{l_1^{5/2}} \label{eq:size_salt_high2}
\end{eqnarray}
and 
\begin{eqnarray}
       w_{eff} & = &  w + \frac{4\pi w_{cc}}{\kappa^2}, \label{eq:w_eff_high2}
\end{eqnarray}
respectively. 

Hence, the polyzwitterionic chain behaves like a neutral chain in the presence of a large amount of salt 
in this repulsion dominated regime and 
the equilibrium radius of gyration of the chain is given by $R_g \sim w_{eff}^{1/5}N^{3/5}$.
 
\section{Conclusions} \label{sec:conclusions}
In summary, we have derived relations for the radius of gyration of a polyzwitterionic chain 
under various solution conditions using the variational 
method. In a salt-free environment, it was shown that the dipole-dipole interactions always lead to the shrinkage 
in the radius of gyration of a purely dipolar polyzwitterionic chain (cf. Eqs. (~\ref{eq:size_salt_free})- (~\ref{eq:nu})). Increase in the dipole moments of the 
zwitterionic monomers or Bjerrum's length was shown to strengthen the attractive dipolar interactions and lead to a stronger chain shrinkage. 
On the addition of salt or in the presence of counterions from the polyzwitterionic chain, the attractive 
charge-dipole and dipole-dipole interactions are screened 
due to the ionic environment created by the the small ions (cf. Eqs. (~\ref{eq:cd_potential}) - (~\ref{eq:dd_potential})). 
The screening of the dipolar interactions is shown to drive the chain 
conformations from the collapsed state to a more open conformational state. These theoretical predictions are  
consistent with the conjecture that the addition 
of salt enhances the solubility\cite{mccormick_review02,bendejacq07} of the polyzwitterionic molecules in aqueous solutions. 

The equilibrium radius of gyration of the polyzwiterionic chain in the presence of the salt 
(Eq. (~\ref{eq:chain_size})) is shown to be an interplay of the relative weightages of screened charge-charge, freely rotating charge-dipole and the dipole-dipole interactions (given by Eqs. (~\ref{eq:wcc}), (~\ref{eq:wcd}) and (~\ref{eq:wdd}), respectively). The weightages depend on the 
degree of counterion adsorption on the zwitterionic sites, the dipole moments of the 
zwitterionic sites, and the ion-pairs formed due to the counterion adsorption. In particular, the equilibrium radius 
of gyration of the polyzwitterionic chain 
depends sensitively on the asymmetry in counterion adsorption on the 
zwitterionic sites. If the counterion adsorption asymmetry ($|\alpha_{+} - \alpha_{-}|$) is small, 
the attractive dipolar interactions leads to the collapse of the chain and the addition 
of the salt leads to the opening of the collapsed globule (see Eq. (~\ref{eq:size_salt_attraction})). On the other hand, if the counterion adsorption 
asymmetry is large, the charge-charge repulsive interactions 
among similarly charged sites dominate over the attractive charge-charge interactions among oppositely charged sites, 
charge-dipole and the dipole-dipole interactions. In this scenario, the polyzwitterionic chain behaves like a 
polyelectrolyte chain (cf. Eqs. (~\ref{eq:size_salt_repulsion}) and (~\ref{eq:size_salt_high2})) with renormalized charge ($ = e|\alpha_{+} - \alpha_{-}|N$). Cross-over from the attraction 
dominated regime for low counterion adsorption asymmetry to the polyelectrolyte regime requires a minimum 
amount of salt, which is consistent with the experimental requirement of a critical amount of salt to solubilize 
the polyzwitterions. Note that these effects are all in addition to the 
conventional role of solvent quality on the conformational characteristics of a polymer chain. 

Furthermore, it is shown that the degree of ionization of the polyzwitterionic chain depends 
on the specificity of the added salt and the zwitterionic groups (cf. Fig. ~\ref{fig:asymmetry}). This explains the specificity of the added salt 
and the zwitterionic groups in determining the conformational behavior of the polyzwitterionic chain. 

Overall, the theoretical predictions are qualitatively consistent with the observed experimental results. 
\section*{ACKNOWLEDGEMENT}
\setcounter {equation} {0} \label{acknowledgement}
 We acknowledge
financial support from the Institute for Collaborative Biotechnologies, and the National Science 
Foundation CMMT  Program under Award DMR-0603710 at the University of
California, Santa Barbara.

\renewcommand{\theequation}{A-\arabic{equation}}
  \setcounter{equation}{0}  
  \section*{APPENDIX A : Screening in dipolar media } \label{app:A}
\begin{figure}[h]
 \begin{center}
     \vspace*{1.0cm}
      \begin{minipage}[c]{10cm}
     \includegraphics[width=10cm, height=5cm]{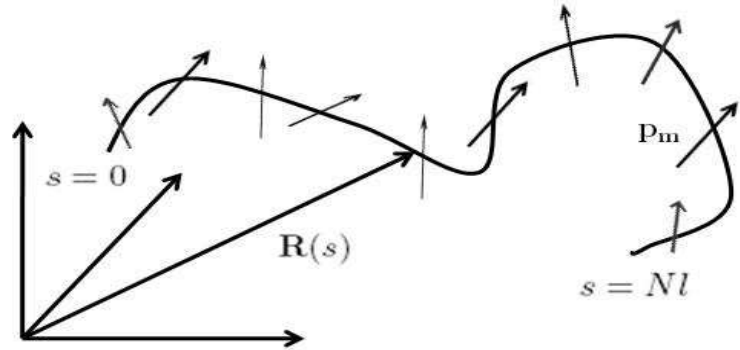}
    \end{minipage}
 \end{center}
\caption{Cartoon of a polyzwitterionic chain modelled as a continuous curve of length 
$Nl$, $l$ being the Kuhn segment length. An arc length variable $s$ is used to represent any point 
along the curve. The polyzwitterionic sites on the side groups in a real chain are modelled 
as dipoles (each with dipole moment of $\mathbf{p}_m$) oriented at some angle to the continuous curve.  } \label{fig:cartoon}
\end{figure}

\begin{figure}[ht] \centering
             \includegraphics[width=6.5in, height=7.5in]{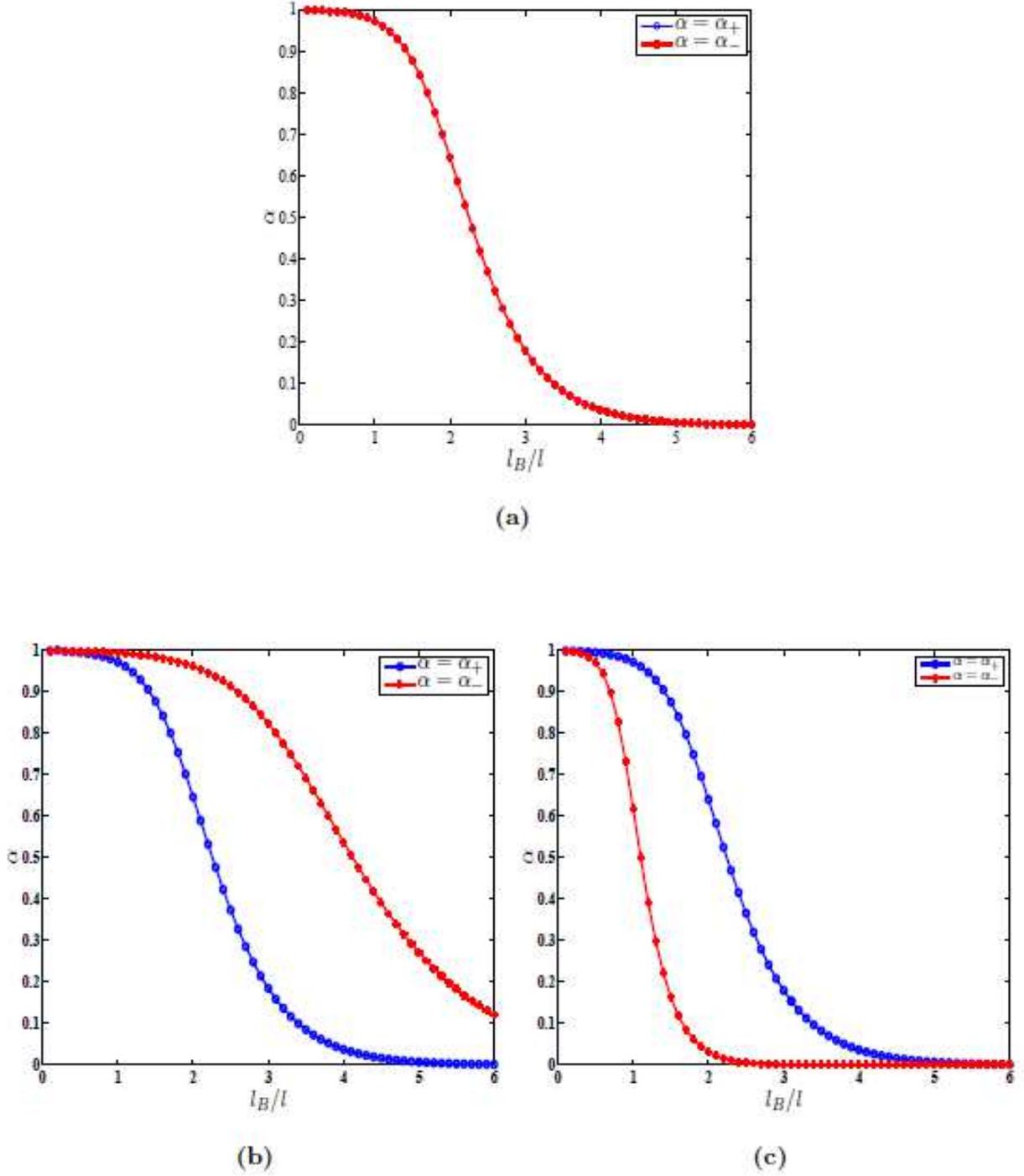}
 \caption{Dependence of the asymmetry in the degree of counterion adsorption ($= |\alpha_{+} - \alpha_-|$) on the specificity of the salt is demonstrated in this figure. Figures (a), (b) and (c) correspond to $\delta_{+} = \delta_{-} = 3.5$; 
 $\delta_{+} = 3.5, \delta_{-} = 2.0$ and $\delta_{+} = 3.5, \delta_{-} = 7.0$, respectively. These figures 
 are obtained for $c_s = 1 mM, N = 1000, Nl^3/\Omega = 0.001, p_m/el = r_d/l = 1$, and $p_{\pm}/el = 0.1$.} \label{fig:asymmetry}
\end{figure}

Here, we present the details of the free energy calculations for a single flexible
polyzwitterionic chain in the presence of salt ions represented in Fig. ~\ref{fig:cartoon}. 
In order to simplify the analysis, consider
a polyzwitterionic chain, where each charged group on the zwitterionic side group is ionized so that the counterion adsorption is absent (i.e., $\alpha_+ = \alpha_- = 1$). We start from the partition function written using an extension of the Edward's Hamiltonian as
\begin{eqnarray}
       Z & = & \frac {1}{\prod_{j}n_{j}!}\int D[\mathbf{R}] \int \prod_{p=1}^{N} d\mathbf{u}_{p} \int \prod_{j} \prod_{m=1}^{n_{j}} d\mathbf{r}_{m} \exp \left [-H_0\left\{\mathbf{R}\right\} - H_w\left\{\mathbf{R},\mathbf{u}_p,\mathbf{R}',\mathbf{u}_p'\right\} \right.\nonumber \\
&& \left.  - H_{cp}\left\{\mathbf{R},\mathbf{u}_{p},\mathbf{r}_m\right\} - H_{pp}\left\{\mathbf{R},\mathbf{u}_{p},\mathbf{R}',\mathbf{u}_{p}'\right\}  - H_{cc}\left\{\mathbf{r}_m,\mathbf{r}_m'\right\} \right ] \label{eq:parti_sing}
\end{eqnarray}
so that
\begin{eqnarray}
      H_0\left\{\mathbf{R}\right\} & = & \frac {3}{2 l}\int_{0}^{Nl} ds\left(\frac{\partial \mathbf{R}(s)}{\partial s} \right )^{2} \label{eq:connectivity}\\
H_w\left\{\mathbf{R},\mathbf{u}_p,\mathbf{R}',\mathbf{u}_p'\right\} &=& \frac{1}{2}\int d\mathbf{r} \int d\mathbf{r}'\int d\mathbf{u} \int d\mathbf{u}' \hat{\rho}_{p}(\mathbf{r},\mathbf{u})w_2(\mathbf{r},\mathbf{u},\mathbf{r}',\mathbf{u}')\hat{\rho}_{p}(\mathbf{r}',\mathbf{u}') \nonumber \\
&& \label{eq:dispersion}\\
H_{pp}\left\{\mathbf{R},\mathbf{u}_{p},\mathbf{R}',\mathbf{u}_p'\right\} &=& \frac{l_B}{2}\int d\mathbf{r}\int d\mathbf{r}' \left[ \frac{Z_{+}^2\hat{\rho}_{m}(\mathbf{r} )\hat{\rho}_{m}(\mathbf{r}' )}{|\mathbf{r} - \mathbf{r}' + 0.5r_d\mathbf{u}_p -0.5r_d\mathbf{u}_p'|} \right . \nonumber \\
&& + \frac{Z_{-}^2\hat{\rho}_{m}(\mathbf{r})\hat{\rho}_{m}(\mathbf{r}')}{|\mathbf{r} - \mathbf{r}' - 0.5r_d\mathbf{u}_p + 0.5r_d\mathbf{u}_p'|} + \frac{Z_{+}Z_{-}\hat{\rho}_{m}(\mathbf{r} )\hat{\rho}_{m}(\mathbf{r}')}{|\mathbf{r} - \mathbf{r}' + 0.5r_d\mathbf{u}_p + 0.5r_d\mathbf{u}_p'|} \nonumber \\
&& \left . +  \frac{Z_{+}Z_{-}\hat{\rho}_{m}(\mathbf{r}' )\hat{\rho}_{m}(\mathbf{r} )}{|\mathbf{r} - \mathbf{r}' - 0.5r_d\mathbf{u}_p - 0.5r_d\mathbf{u}_p'|} \right ]\label{eq:segelec}\\
&=&  \frac{l_B}{2}\int d\mathbf{r} \int d\mathbf{r}'\int d\mathbf{u} \int d\mathbf{u}' \hat{\rho}_{p}(\mathbf{r},\mathbf{u})w_{pp}(\mathbf{r},\mathbf{u},\mathbf{r}',\mathbf{u}')\hat{\rho}_{p}(\mathbf{r}',\mathbf{u}') \nonumber \\
&& \label{eq:segdipole}\\
H_{cp}\left\{\mathbf{R},\mathbf{u}_{p}',\mathbf{r}_m\right\} &=&  l_B\int d\mathbf{r}\int d\mathbf{r}' \left(\sum_j Z_j\hat{\rho}_{j}(\mathbf{r})\right)\nonumber \\
&& \quad \left(\frac{Z_{+}\hat{\rho}_{m}(\mathbf{r}' )}{|\mathbf{r} - \mathbf{r}' - 0.5r_d\mathbf{u}_p'|}  + \frac{Z_{-}\hat{\rho}_{m}(\mathbf{r}' )}{|\mathbf{r} - \mathbf{r}' + 0.5r_d\mathbf{u}_p'|}\right), \label{eq:countseg}\\
 H_{cc}\left\{\mathbf{r}_m,\mathbf{r}_m'\right\} & = & \frac{l_B}{2}\int d\mathbf{r} \int d\mathbf{r}'\frac{\left(\sum_j Z_j\hat{\rho}_{j}(\mathbf{r})\right)\left(\sum_j Z_j\hat{\rho}_{j}(\mathbf{r}')\right)}{|\mathbf{r} - \mathbf{r}' |},\label{eq:countcount}
\end{eqnarray}
where $\mathbf{R}(s)$ represents the position vector for the $s^{th}$ segment (see Fig. ~\ref{fig:cartoon}) and subscripts $j = c+,c-,s+,s-$ represent the small ions released by the polyzwitterionic chain along with those coming from the added salt. 
In Eq. (~\ref{eq:parti_sing}), the Hamiltonian is written by taking into account 
the contributions coming from the chain connectivity (given by $H_0$ in Eq. (~\ref{eq:connectivity})), the short ranged dispersion interactions (represented by $H_w$ 
in Eq. (~\ref{eq:dispersion})) and the long range electrostatic 
interactions between the charged species (written as $H_{pp}, H_{cp}$ and $H_{cc}$ above, which correspond to the segment-segment, segment-small ions and small ions-small ions 
interactions, respectively).  

The chain connectivity part written as $H_0$ is the well-known Wiener measure for a 
flexible polymer chain. Furthermore, $w_{2}(\mathbf{r},\mathbf{u},\mathbf{r}\,',\mathbf{u}\,')$ is the energy accounting for dispersion 
interactions between segments having centers at $\mathbf{r}$ and $\mathbf{r}\,'$, and their axes along the direction $\mathbf{u}$ and $\mathbf{u}\,'$, respectively. For the 
large length scale properties such as the radius of gyration described within the coarse-grained model, functional form for the dispersion interactions doesn't matter\cite{edwardsbook}. What matters is the short range nature of these 
interactions. 

Electrostatic contributions to the Hamiltonian arising from the segment-segment  interactions can be written by taking into account the Coulomb interactions between the charges at the ends of the dipoles located at each segment. Adding the Coulomb interaction energies between the charges at the ends of the two dipoles 
of length $r_d$ with their centers at $\mathbf{R}(s)$ and $\mathbf{R}(s\,')$, and the dipolar axes along the directions described by unit vectors $\mathbf{u}_p(s)$ and $\mathbf{u}_p(s\,')$, respectively, and summing over all the dipoles, Eq. (~\ref{eq:segelec}) is readily obtained. 
In writing Eq. (~\ref{eq:segelec}), the microscopic number density of the dipoles (or segments ) at a certain location $\mathbf{r}$ is defined as 
        \begin{eqnarray}
\hat{\rho}_{m}(\mathbf{r})  &=& \frac{1}{l}\int_{0}^{Nl} dt \, \delta \left[\mathbf{r}-\mathbf{R}(s)\right]
\end{eqnarray}
and the functional dependence of vector $\mathbf{u}_p(s)$ on $s$ has been suppressed for the ease in writing. 

A simpler way to rewrite the segment-segment interaction energy is by defining a distribution function ($\hat{\rho}_p(\mathbf{r},\mathbf{u})$), which describes the number of segments with their centers at a certain location $\mathbf{r}$ and the attached 
dipoles oriented along $\mathbf{u}$. If the dipole attached to the segment at $\mathbf{R}(s)$ has its dipolar axis along $\mathbf{u}_p(s)$, then the distribution function can 
be formally defined as  
        \begin{eqnarray}
    \hat{\rho}_{p}(\mathbf{r},\mathbf{u})  &=& \frac{1}{l}\int_{0}^{Nl} ds \, \delta \left[\mathbf{r}-\mathbf{R}(s)\right]\delta \left[\mathbf{u}-\mathbf{u}_p(s)\right].
\end{eqnarray}
Also, if $w_{pp}(\mathbf{r},\mathbf{u},\mathbf{r}\,',\mathbf{u}\,')$ is the 
interaction energy between the dipoles with their centers at $\mathbf{r}$ and $\mathbf{r}\,'$, and their dipolar axis along the direction $\mathbf{u}$ and
$\mathbf{u}\,'$, respectively, then the segment-segment interaction energy is given by Eq. (~\ref{eq:segdipole}). It can be readily shown that the Eqs. (~\ref{eq:segelec}) and (~\ref{eq:segdipole}) are equivalent. In fact, an explicit expression for the interaction 
energy arising from the long range dipole-dipole interactions i.e., $w_{pp}(\mathbf{r},\mathbf{u},\mathbf{r}\,',\mathbf{u}\,')$, can be derived using 
the Taylor expansion for the function $1/|\mathbf{r} - \mathbf{p}|$ in the limit of $|\mathbf{r}| \gg |\mathbf{p}|$. However, we leave the functional form for  
$w_{pp}(\mathbf{r},\mathbf{u},\mathbf{r}\,',\mathbf{u}\,')$ for now and first focus on 
carrying out integrations over the positions of the small ions in Eq. (~\ref{eq:parti_sing}).

Like the segment-segment interaction energies described by the dipole-dipole interactions , interactions between the small ions and the segments, and among the small ions are described by the charge-dipole and charge-charge interactions given in Eqs. (~\ref{eq:countseg}) and (~\ref{eq:countcount}), respectively. Note that in writing the interaction energies involving the small ions (in Eqs. (~\ref{eq:countseg}) and (~\ref{eq:countcount})),
we have taken the small ions to be point charges so that
they have zero excluded volume, and hence, interactions are purely electrostatic in nature.
Furthermore, the microscopic densities for the small ions are defined as
        \begin{eqnarray}
     \hat{\rho}_{j}(\mathbf{r}) &=&  \sum_{i=1}^{n_{j}} \delta \left[\mathbf{r}-\mathbf{r}_i)\right] \quad \mbox{for} \quad j = c+,c-,s+,s-. 
\end{eqnarray}
In order to carry out the integrations over the positions of the small ions, we use the identity
        \begin{eqnarray}
 1 &=& \int \prod_j D\left[w_j\right]\int \prod_j D\left[\rho_j\right] \exp \left[i\int d\mathbf{r} \sum_j w_j(\mathbf{r})(\rho_j(\mathbf{r}) - \hat{\rho}_j(\mathbf{r}))\right]
\end{eqnarray}
for $j = c+,c-,s+,s-$. Using this identity and Stirling's approximation $\ln n! \simeq n\ln n - n$, Eq. (~\ref{eq:parti_sing}) becomes
\begin{eqnarray}
       Z & = & \int D[\mathbf{R}] \int \prod_{p=1}^{N} d\mathbf{u}_{p} \int \prod_{j} D\left[w_j\right]\int \prod_j D\left[\rho_j\right] \exp \left [-H\left\{\mathbf{R},\mathbf{u}_p,w_j,\rho_j\right\}\right ]
\end{eqnarray}
where
\begin{eqnarray}
       H\left\{\mathbf{R},\mathbf{u}_p,w_j,\rho_j\right\} & = & H_1\left\{\mathbf{R},\mathbf{u}_p\right\} + H_2\left\{\mathbf{R},\mathbf{u}_p,w_j,\rho_j\right\}.
\end{eqnarray}
Here, $H_1$ and $H_2$ represent the parts of the Hamiltonian, which are independent and dependent on $w_j$, respectively.
Now, evaluating functional integrals over $w_j$ by the value of the integrand at the saddle points $i.e., \int \prod_{j} D\left[w_j\right] \exp \left [-H_2\left\{\mathbf{R},\mathbf{u}_p,w_j,\rho_j\right\}\right ]\simeq \exp \left [-H_2\left\{\mathbf{R},\mathbf{u}_p,w_j^\star,\rho_j\right\}\right ]$, partition
function becomes
\begin{eqnarray}
       Z & = & \int D[\mathbf{R}] \int \prod_{p=1}^{N} d\mathbf{u}_{p} \int \prod_j D\left[\rho_j\right] \exp \left [-H\left\{\mathbf{R},\mathbf{u}_p,w_j^\star,\rho_j\right\}\right ].
\end{eqnarray}
Here, $w_j^\star$ is obtained by using $\frac{\delta H_2}{\delta w_j}|_{w_j=w_j^\star} = 0$. Explicitly, $H_2\left\{\mathbf{R},\mathbf{u}_p,w_j^\star,\rho_j\right\}$ is given by

\begin{eqnarray}
       H_2\left\{\mathbf{R},\mathbf{u}_p,w_j^\star,\rho_j\right\} &=& \frac{l_B}{2}\int d\mathbf{r} \int d\mathbf{r}'\left[\frac{\left(\sum_j Z_j\rho_{j}(\mathbf{r})\right)\left(\sum_j Z_j\rho_{j}(\mathbf{r}')\right)}{|\mathbf{r} - \mathbf{r}' |} \right . \nonumber \\
&& \left. + l_B \left(\sum_j Z_j\rho_{j}(\mathbf{r})\right)\left(\frac{Z_{+}\hat{\rho}_{m}(\mathbf{r}')}{|\mathbf{r} - \mathbf{r}' - 0.5r_d\mathbf{u}_p'|}  + \frac{Z_{-}\hat{\rho}_{m}(\mathbf{r}')}{|\mathbf{r} - \mathbf{r}' + 0.5r_d\mathbf{u}_p'|}\right)\right] \nonumber \\
&&  +  \int d\mathbf{r} \sum_j \rho_j(\mathbf{r})(\ln \rho_j(\mathbf{r}) - 1).
\end{eqnarray}
 To evaluate the functional integrals over $\rho_j$'s, we use the well-known random phase approximation, which amounts to the expansion
$\rho_{j}(\mathbf{r}) = \bar{\rho}_{j} + \delta \rho_{j}(\mathbf{r})$, under the constraint $\int d\mathbf{r} \delta \rho_{j}(\mathbf{r}) = 0$. The constraint implies $\bar{\rho}_{j} = n_j/\Omega$. Now, expressing $H_2$ in powers of $\delta \rho_{j}(\mathbf{r})$ and retaining up to quadratic terms,
the integrals over $\delta \rho_{j}(\mathbf{r})$ can be carried out analytically. The result is
\begin{eqnarray}
       Z & = & \Gamma Z_0\int D[\mathbf{R}] \int \prod_{p=1}^{N} d\mathbf{u}_{p} \exp \left [-H_1\left\{\mathbf{R},\mathbf{u}_p\right\} + \frac{\kappa^2}{8\pi l_B}  \int \frac{d^3 \mathbf{k}}{(2\pi)^3} \frac{V(k,\mathbf{u}_p)V(-k,\mathbf{u}_p')}{1 + \frac{\kappa^2}{k^2} }\right ],\nonumber \\
       &&
\end{eqnarray}
where
\begin{eqnarray}
-\ln Z_{0} &=& \Omega \bar{\rho}_j (\ln \bar{\rho}_j - 1) + \frac{\Omega}{2} \int \frac{d^3 \mathbf{k}}{(2\pi)^3} \ln \left(1 + \frac{\kappa^2}{k^2}\right ).
\end{eqnarray}
Here, $\kappa^2 = 4\pi l_B \sum_j Z_j^2 \bar{\rho}_j = 4\pi l_B \sum_j Z_j^2 n_j/\Omega$ and $\Gamma$ is the normalizing factor. Also, $V(k,u_p')$
is given by
\begin{eqnarray}
       V(k,\mathbf{u}_p') &=& \int d \mathbf{r}  V(\mathbf{r},\mathbf{u}_p')e^{i\mathbf{k}.\mathbf{r}}\nonumber \\
&=& l_B \int d \mathbf{r} \int d \mathbf{r}' \left(\frac{Z_{+}\hat{\rho}_{m}(\mathbf{r}')}{|\mathbf{r} - \mathbf{r}' - 0.5r_d\mathbf{u}_p'|}  + \frac{Z_{-}\hat{\rho}_{m}(\mathbf{r}')}{|\mathbf{r} - \mathbf{r}' + 0.5r_d\mathbf{u}_p'|}\right)  e^{i\mathbf{k}.\mathbf{r}} \nonumber \\
&=& \frac{4\pi l_B}{k^2} \left[Z_{+}\hat{\rho}_{m}(k)e^{0.5r_d i \mathbf{k}.\mathbf{u}_p'} + Z_{-}\hat{\rho}_{m}(k)e^{-0.5r_d i \mathbf{k}.\mathbf{u}_p'}\right].
\end{eqnarray}
Using this expression for $V(k,\mathbf{u}_p')$, the partition function becomes
\begin{eqnarray}
       Z & = & \Gamma Z_0\int D[\mathbf{R}] \int \prod_{p=1}^{N} d\mathbf{u}_{p} \exp \left [-H_0\left\{\mathbf{R}\right\} - H_w\left\{\mathbf{R},\mathbf{u}_p,\mathbf{R}',\mathbf{u}_p'\right\} - H_{pp}'\left\{\mathbf{R},\mathbf{u}_{p},\mathbf{R}',\mathbf{u}_{p}'\right\} \right ]
\nonumber \\
&& \label{eq:parti_sing_new}
\end{eqnarray}

where
\begin{eqnarray}
H_{pp}'\left\{\mathbf{R},\mathbf{u}_{p},\mathbf{R}',\mathbf{u}_p'\right\} &=& \frac{l_B}{2}\int d\mathbf{r}\int d\mathbf{r}' \left[
\frac{Z_{+}^2\hat{\rho}_{m}(\mathbf{r})\hat{\rho}_{m}(\mathbf{r}')e^{-\kappa|\mathbf{r} - \mathbf{r}' + 0.5r_d\mathbf{u}_p -0.5r_d\mathbf{u}_p'|}}{|\mathbf{r} - \mathbf{r}' + 0.5r_d\mathbf{u}_p -0.5r_d\mathbf{u}_p'|} \right . \nonumber \\
&& + \frac{Z_{-}^2\hat{\rho}_{m}(\mathbf{r})\hat{\rho}_{m}(\mathbf{r}')e^{-\kappa |\mathbf{r} - \mathbf{r}' - 0.5r_d\mathbf{u}_p
+ 0.5r_d\mathbf{u}_p'|}}{|\mathbf{r} - \mathbf{r}' - 0.5r_d\mathbf{u}_p + 0.5r_d\mathbf{u}_p'|} \nonumber \\
&& + \frac{Z_{+}Z_{-}\hat{\rho}_{m}(\mathbf{r})\hat{\rho}_{m}(\mathbf{r}')e^{-\kappa |\mathbf{r} - \mathbf{r}' + 0.5r_d\mathbf{u}_p + 0.5r_d\mathbf{u}_p'|}}{|\mathbf{r} - \mathbf{r}' + 0.5r_d\mathbf{u}_p + 0.5r_d\mathbf{u}_p'|} \nonumber \\
&& \left . +  \frac{Z_{+}Z_{-}\hat{\rho}_{m}(\mathbf{r}')\hat{\rho}_{m}(\mathbf{r})e^{-\kappa |\mathbf{r} - \mathbf{r}' - 0.5r_d\mathbf{u}_p - 0.5r_d\mathbf{u}_p'|}}{|\mathbf{r} - \mathbf{r}' - 0.5r_d\mathbf{u}_p - 0.5r_d\mathbf{u}_p'|} \right ]. \label{eq:screened_mm}
\end{eqnarray}
To make some progress analytically, we use the multipole expansion for the function $e^{-\kappa |\mathbf{r} - \mathbf{p}|}/|\mathbf{r} - \mathbf{p}|$. Using the vector Taylor expansion\cite{weber_book} for an arbitrary function $f$
\begin{eqnarray}
f (\mathbf{r} + \mathbf{a}) &=& \sum_{n=0}^{\infty} \frac{1}{n!} (\mathbf{a}.\nabla)^n f (\mathbf{r})
\end{eqnarray}
in the limit of $|\mathbf{r}| \gg |\mathbf{p}|$
\begin{eqnarray}
\frac{e^{-\kappa |\mathbf{r} - \mathbf{p}|}}{|\mathbf{r} - \mathbf{p}|} &=& \frac{e^{-\kappa |\mathbf{r}|}}{|\mathbf{r}|}  + A(\mathbf{p}.\mathbf{r}) - \frac{1}{2}\left[A(\mathbf{p}.\mathbf{p}) - B (\mathbf{p}.\mathbf{r})(\mathbf{p}.\mathbf{r})\right] + O(|\mathbf{p}|^3), \label{eq:screened_mm_2}
\end{eqnarray}
where
\begin{eqnarray}
A  &=& e^{-\kappa |\mathbf{r}|} \left[\frac{1}{|\mathbf{r}|^3} + \frac{\kappa}{|\mathbf{r}|^2} \right ], \\
B &=& e^{-\kappa |\mathbf{r}|} \left[ \frac{3}{|\mathbf{r}|^5} + \frac{3\kappa}{|\mathbf{r}|^4} + \frac{\kappa^2}{|\mathbf{r}|^3}\right ].
\end{eqnarray}
For the purely dipolar case, $Z_+ = -Z_-$, and this causes the monopole and the dipole
terms to disappear when the multipole expansion (Eq. (~\ref{eq:screened_mm_2})) is used in Eq. (~\ref{eq:screened_mm}). The lowest order term that survive is
the quadrupole term, which is reponsible for the dipole-dipole interactions. Explicitly,

\begin{eqnarray}
H_{pp}'\left\{\mathbf{R},\mathbf{u}_{p},\mathbf{R}',\mathbf{u}_p'\right\} &=& \frac{l_B}{2e^2}\int d\mathbf{r}\int d\mathbf{r}' \hat{\rho}_{m}(\mathbf{r})\hat{\rho}_{m}(\mathbf{r}') \left[ A(\mathbf{p}_m.\mathbf{p}_m') - B \left[\mathbf{p}_m.(\mathbf{r} - \mathbf{r}')\right ]\left[\mathbf{p}_m'.(\mathbf{r} - \mathbf{r}')\right]  \right] \nonumber \\
&=&  \frac{1}{2}\int d\mathbf{r} \int d\mathbf{r}'\int d\mathbf{u} \int d\mathbf{u}' \hat{\rho}_{p}(\mathbf{r},\mathbf{u})w_{pp}^s(\mathbf{r},\mathbf{u},\mathbf{r}',\mathbf{u}')\hat{\rho}_{p}(\mathbf{r}',\mathbf{u}'),
\end{eqnarray}
where $\mathbf{p}_m = r_d Z_{+}e\mathbf{u}_p$ and $\mathbf{p}_m' = r_d Z_{+}e\mathbf{u}_p'$ are the dipole moments of the zwitterionic side groups. Physically, this means that the charged moieties on the zwitterionic groups interact with each other by a
screened dipole-dipole interaction potential given by
\begin{eqnarray}
w_{pp}^s(\mathbf{r},\mathbf{p}_m,\mathbf{r}',\mathbf{p}_m') &=& \frac{l_B}{e^2}\left[ A(\mathbf{p}_m.\mathbf{p}_m') - B \left[\mathbf{p}_m.(\mathbf{r} - \mathbf{r}')\right ]\left[\mathbf{p}_m'.(\mathbf{r} - \mathbf{r}')\right]  \right],
\end{eqnarray}
where $\mathbf{r}$ is to be replaced by $\mathbf{r} - \mathbf{r}'$ in expressions for $A$ and $B$ above. Furthermore, note that putting $\kappa = 0$ in the expressions for $A$ and 
$B$, well-known functional form the bare dipole-dipole interaction energy\cite{intermolecular_forces} represented by  $w_{pp}(\mathbf{r},\mathbf{u},\mathbf{r}',\mathbf{u}')$ in Eq. (~\ref{eq:segdipole}) can be readily derived. 

So far, we have carried out the integrations over the positions of the small ions in the partition function. Now, we need to carry out the integrations over the orientations of the dipoles attached to the segments (i.e, the integrals over $\mathbf{u}_{p}$ in Eq. (~\ref{eq:parti_sing_new})). In order to carry out these integrals, we assume that the strength of dipole-dipole interactions is weak and hence, the dipoles can rotate freely with respect to each other. For the freely rotating dipoles\cite{intermolecular_forces}, integrals over $\mathbf{u}_p$ in Eq. (~\ref{eq:parti_sing_new}) can be carried out by expanding the exponential in powers of $\mathbf{u}_p$  and $\mathbf{u}_p'$
(i.e., $\mathbf{p}_m$ and $\mathbf{p}_m'$) and exponentiating the series after carrying out the integrals. Expansion
up to second degree terms give
\begin{eqnarray}
I_{dd} &=& \int d\mathbf{u}_p \int d\mathbf{u}_p'  \exp \left [ - \frac{l_B}{2e^2}\left \{ A(\mathbf{p}_m.\mathbf{p}_m') - B \left[\mathbf{p}_m.(\mathbf{r} - \mathbf{r}')\right ]\left[\mathbf{p}_m'.(\mathbf{r} - \mathbf{r}')\right ]\right\} \right ]\nonumber \\
&=& \frac{1}{(4\pi)^2}\int_{0}^{\pi} d\theta_1 \sin \theta_1 \int_{0}^{2\pi} d\phi_1 \int_{0}^{\pi} d\theta_2 \sin \theta_2 \int_{0}^{2\pi} d\phi_2 \nonumber \\
&& \exp \left [ - \frac{l_B}{2e^2}\left \{ A(\mathbf{p}_m.\mathbf{p}_m') - B \left[\mathbf{p}_m.(\mathbf{r} - \mathbf{r}')\right ]\left[\mathbf{p}_m'.(\mathbf{r} - \mathbf{r}')\right ]\right\} \right ]  \nonumber \\
&\simeq&  \exp \left [ \frac{1}{12}  \left(\frac{p_m p_m'}{e^2}\right)^2 l_B^2 \frac{\exp \left[-2\kappa |\mathbf{r} - \mathbf{r}'|\right]}{|\mathbf{r} - \mathbf{r}'|^6} C(\kappa|\mathbf{r} - \mathbf{r}'|)\right ], \label{eq:dipole_dipole}
\end{eqnarray}
where
\begin{eqnarray}
C(x) &=& 1 + 2x + \frac{5}{3}x^2 + \frac{2}{3}x^3 + \frac{1}{6}x^4. \label{eq:c_equation}
\end{eqnarray}

This completes the treatment of the electrostatic terms in Eq. (~\ref{eq:parti_sing}). 
Now, consider the general case, where some of the counterions from the solution may 
adsorb on the zwitterionic monomers and a small dipole (of dipole moment $\mathbf{p}_+$ or $\mathbf{p}_-$ corresponding to positive or negative counterion adsorption, respectively ) is formed at the adsorption site. 
In this particular situation, one has to take into account the distribution function 
of the adsorbed ions. However, the counterions may adsorb and desorb frequently as a result of the thermal fluctuations. In other words, on 
an average, the charge on the backbone is smeared rather than localized. 
Furthermore, one has to take into account the adsorption energy and the entropy of the distribution of the counterions among the zwitterionic sites on the chain. That's why we consider the so called ``permuted'' charge distribution\cite{rajeev_arindam} so that the probability of finding a bare positive or negative charge on the polyzwitterionic chain is $\alpha_+$ or $\alpha_-$, respectively. Dividing the whole population of small ions on the basis of ``free'' and ``adsorbed'' states, we rewrite the Hamiltonian with the probabilities of finding charges and dipoles on the chain. Now, carrying out the same analysis as presented above, it is found that the charge-charge and charge-dipole 
terms also survive in the partition function with suitable prefactors characterizing the weightage of different kinds of interactions (charge-charge, charge-dipole and dipole-dipole). These prefactors are given by $w_{cc}, w_{cd}$ and $w_{dd}$ in Eqs. (~\ref{eq:wcc}) - (~\ref{eq:wdd}) for the charge-charge, charge-dipole and dipole-dipole interactions, respectively. Extra contributions coming from the charge-dipole interactions require the evaluation of the integrals over the orientations of the dipoles given by
\begin{eqnarray}
I_{cd} &=& \int d\mathbf{u}_p \exp \left [ - \frac{l_B Q }{e^2}A(\mathbf{p}.(\mathbf{r} - \mathbf{r}')) \right ]\nonumber \\
&=& \frac{1}{(4\pi)}\int_{0}^{\pi} d\theta_1 \sin \theta_1 \int_{0}^{2\pi} d\phi_1 \exp \left [ - \frac{l_B Q }{e^2}A(\mathbf{p}.(\mathbf{r} - \mathbf{r}')) \right ]  \nonumber \\
&\simeq&  \exp \left [ \frac{1}{6} Q^2 \left(\frac{p}{e}\right)^2 l_B^2 \frac{\exp \left[-2\kappa |\mathbf{r} - \mathbf{r}'|\right]}{|\mathbf{r} - \mathbf{r}'|^4} (1 + \kappa|\mathbf{r} - \mathbf{r}'|)^2\right ]. \label{charge_dipole}
\end{eqnarray}
Physically, the integrals involve the interactions between a charge $Q$ (in terms of electronic charge) and a dipole of moment $p$. 

In writing Eq. (~\ref{eq:partition_main}), we take the angular average of the short range part in the Hamiltonian (i.e., $H_w$) to define an excluded volume parameter $w$. Also, we add the repulsive ternary interactions characterized by $\nu > 0$ to stabilize against the attractive binary interactions.

\renewcommand{\theequation}{B-\arabic{equation}}
  \setcounter{equation}{0}  
  \section*{APPENDIX B : Uniform expansion model : effect of dipolar interactions} \label{app:B}

Carrying out integrals over $s,s'$ and $k$ in Eqs. (~\ref{eq:icd}) and (~\ref{eq:idd}) 
\begin{eqnarray}
I_{cd} &=& \frac{4}{3}\left(\frac{3}{2\pi}\right)^{3/2} \frac{(Nl)^{1/2}}{l_1^{5/2}}\frac{3\pi}{\lambda} 
+ \frac{3}{l_1^3}\left \{ \frac{2\sqrt{b}}{\sqrt{\pi}} - \frac{5}{b} -  4 + \frac{10}{\sqrt{\pi}\sqrt{b}} - \frac{b-5}{b}\exp \left(b\right)\mbox{erfc}(\sqrt{b})\right \} \nonumber \\
&& - \frac{12}{l_1^3} \int_{0}^{1}ds \int_{0}^{s}ds' \frac{\exp \left[b (s-s')\right]}{s-s'}\mbox{erfc}\left(\sqrt{b(s-s')}\right),
\end{eqnarray}
and
\begin{eqnarray}
I_{dd} &=& \frac{4}{3}\left(\frac{3}{2\pi}\right)^{3/2} \frac{(Nl)^{1/2}}{l_1^{5/2}} \left[4\pi \left(\frac{7}{48 \lambda^3} + \frac{\kappa}{3\lambda^2} + \frac{\kappa^2}{6\lambda} \right )\right ] + 15\sqrt{\frac{6}{\pi}}\frac{\kappa}{(Nl)^{1/2}l_1^{7/2}} \nonumber \\
&& + \frac{\kappa^2}{l_1^3} \left \{ \frac{41}{2b}  + 15  -  \frac{107}{\sqrt{\pi}\sqrt{b}} + \frac{\sqrt{b}}{\sqrt{\pi}} 
 - \frac{41 - 11 b}{2b}\exp \left(b\right)\mbox{erfc}(\sqrt{b})\right \} \nonumber \\
&& + \frac{\kappa^2}{l_1^3}\int_{0}^{1}ds \int_{0}^{s}ds' \exp \left[b(s-s')\right]\mbox{erfc}\left(\sqrt{b (s-s')}\right) \left [ \frac{12}{b (s-s')^2} + \frac{4}{(s-s')}\right ], 
\end{eqnarray}

where $b = 4\kappa^2 N l l_1/6 =  4a$.

Unfortunately, analytical evaluations of the integrals over $s$ and $s'$ in the expressions for $I_{cd}$ and $I_{dd}$ are
not possible for an arbitrary $b$. However, in the limiting cases for $b$, these integrals can be readily evaluated 
using the asymptotic expansion for the function $\exp (x)\mbox{erfc}\left(\sqrt{x}\right)$. Consider the 
case of small $b$ first. 

\subsection{$b\rightarrow 0$}
In this limit, the integrals can be carried out using the asymptotic expansion
\begin{eqnarray}
\exp (x)\mbox{erfc}\left(\sqrt{x}\right)_{|x\rightarrow 0}  &=& 1 - \frac{2 \sqrt{x}}{\sqrt{\pi}}  + x  - \frac{4 x^{3/2}}{3\sqrt{\pi}}
\end{eqnarray}
so that 
\begin{eqnarray}
\frac{12}{l_1^3} \int_{0}^{1}ds \int_{0}^{s}ds' \frac{\exp \left[b (s-s')\right]}{s-s'}\mbox{erfc}\left(\sqrt{b(s-s')}\right) &=& 
\frac{12}{l_1^3} \int_{0}^{1}ds \int_{0}^{s}ds' \frac{1}{s-s'} \nonumber \\
&& - \frac{1}{l_1^3} \left[ \frac{32 \sqrt{b}}{\sqrt{\pi}} - 6 b + \frac{64 b^{3/2}}{15 \sqrt{\pi}} \right]
\end{eqnarray}
and 
\begin{eqnarray}
\frac{\kappa^2}{l_1^{3}}\int_{0}^{1}ds \int_{0}^{s}ds' \exp \left[b(s-s')\right]\mbox{erfc}\left(\sqrt{b (s-s')}\right) \left [ \frac{12}{b (s-s')^2} + \frac{4}{(s-s')}\right ] &=& \nonumber \\
\frac{16 \kappa^2}{l_1^3}\int_{0}^{1}ds \int_{0}^{s}ds' \frac{1}{(s-s')}  + \frac{18}{N l l_1^{4}}\int_{0}^{1}ds \int_{0}^{s}ds' \frac{1}{(s-s')^{2}}&& \nonumber \\
 + \frac{\kappa^2}{l_1^3} \left[\frac{96}{\sqrt{\pi}\sqrt{b}} - \frac{96\sqrt{b}}{\sqrt{\pi}} + 2b 
 -  \frac{64 b^{3/2}}{45 \sqrt{\pi}} \right ].&&
\end{eqnarray}

Assembling the pieces, expressions for $I_{cd}$ and $I_{dd}$ in the limit of small $b$ become 
\begin{eqnarray}
I_{cd} &=& \frac{4}{3}\left(\frac{3}{2\pi}\right)^{3/2} \frac{(Nl)^{1/2}}{l_1^{5/2}}\frac{3\pi}{\lambda} 
- \frac{12}{l_1^3} \int_{0}^{1}ds \int_{0}^{s}ds' \frac{1}{s-s'} \nonumber \\
&+& \frac{1}{l_1^3}\left \{ \frac{38\sqrt{b}}{\sqrt{\pi}} - 6b + \frac{64b^{3/2}}{15\sqrt{\pi}} 
-  3 \left( \frac{5}{b}+ 4 - \frac{10}{\sqrt{\pi}\sqrt{b}} + \frac{b-5}{b}\exp \left(b\right)\mbox{erfc}(\sqrt{b} )\right )\right \} 
\end{eqnarray}
and
\begin{eqnarray}
I_{dd} &=& \frac{4}{3}\left(\frac{3}{2\pi}\right)^{3/2} \frac{(Nl)^{1/2}}{l_1^{5/2}} \left[4\pi \left(\frac{7}{48 \lambda^3} + \frac{\kappa}{3\lambda^2} + \frac{\kappa^2}{6\lambda} \right )\right ] + 15\sqrt{\frac{6}{\pi}}\frac{\kappa}{(Nl)^{1/2}l_1^{7/2}} \nonumber \\
&& + \frac{\kappa^2}{l_1^3} \left \{ \frac{41}{2b}  + 15  -  \frac{11}{\sqrt{\pi}\sqrt{b}} - \frac{31\sqrt{b}}{\sqrt{\pi}} + 2b - \frac{64 b^{3/2}}{45 \sqrt{\pi}} 
 - \frac{41 - 11 b}{2b}\exp \left(b\right)\mbox{erfc}(\sqrt{b})\right \} \nonumber \\
&& + \frac{16 \kappa^2}{l_1^3}\int_{0}^{1}ds \int_{0}^{s}ds' \frac{1}{(s-s')}  + \frac{18}{N l l_1^{4}}\int_{0}^{1}ds \int_{0}^{s}ds' \frac{1}{(s-s')^{2}}.
\end{eqnarray}
Taking the limit of $b\rightarrow 0 $, Eqs. (~\ref{eq:icd_lim}) and (~\ref{eq:idd_lim}) have been obtained.

\subsection{$b\rightarrow \infty$}
In this limit, the integrals can be carried out using the asymptotic expansion
\begin{eqnarray}
\exp (x)\mbox{erfc}\left(\sqrt{x}\right)_{|x\rightarrow \infty}  &=& \frac{1}{\sqrt{\pi x}} 
\end{eqnarray}
so that
\begin{eqnarray}
\frac{12}{l_1^3} \int_{0}^{1}ds \int_{0}^{s}ds' \frac{\exp \left[b (s-s')\right]}{s-s'}\mbox{erfc}\left(\sqrt{b(s-s')}\right) &=&
-\frac{48}{l_1^3\sqrt{\pi} \sqrt{b}} 
\end{eqnarray}
and
\begin{eqnarray}
\frac{\kappa^2}{l_1^{3}}\int_{0}^{1}ds \int_{0}^{s}ds' \exp \left[b(s-s')\right]\mbox{erfc}\left(\sqrt{b (s-s')}\right) \left [ \frac{12}{b (s-s')^2} + \frac{4}{(s-s')}\right ] &=& \nonumber \\
\frac{\kappa^2}{l_1^3}\left[\frac{12}{\sqrt{\pi}b^{3/2}}\int_{0}^{1}ds \int_{0}^{s}ds' \frac{1}{(s-s')^{5/2}}  
- \frac{16}{\sqrt{\pi} \sqrt{b}} \right ]. && 
\end{eqnarray}

Using these integrals, expressions for $I_{cd}$ and $I_{dd}$ in the limit of large $b$ become
\begin{eqnarray}
I_{cd} &=& \frac{4}{3}\left(\frac{3}{2\pi}\right)^{3/2} \frac{(Nl)^{1/2}}{l_1^{5/2}}
\left(\frac{3\pi}{\lambda}\right) \nonumber \\
&& + \frac{3}{l_1^3}\left \{ \frac{2\sqrt{b}}{\sqrt{\pi}} -  \frac{5}{b}- 4 + \frac{26}{\sqrt{\pi}\sqrt{b}} - \frac{b-5}{b}\exp \left(b\right)\mbox{erfc}(\sqrt{b} )\right \}
\end{eqnarray}
and
\begin{eqnarray}
I_{dd} &=& \frac{4}{3}\left(\frac{3}{2\pi}\right)^{3/2} \frac{(Nl)^{1/2}}{l_1^{5/2}} \left[4\pi \left(\frac{7}{48 \lambda^3} + \frac{\kappa}{3\lambda^2} + \frac{\kappa^2}{6\lambda} \right )\right ] + 15\sqrt{\frac{6}{\pi}}\frac{\kappa}{(Nl)^{1/2}l_1^{7/2}} \nonumber \\
&& + \frac{\kappa^2}{l_1^3} \left \{ \frac{41}{2b}  + 15  -  \frac{123}{\sqrt{\pi}\sqrt{b}} + \frac{\sqrt{b}}{\sqrt{\pi}}
 - \frac{41 - 11 b}{2b}\exp \left(b\right)\mbox{erfc}(\sqrt{b})\right \} \nonumber \\
&& + \frac{12\kappa^2}{\sqrt{\pi}l_1^3 b^{3/2}}\int_{0}^{1}ds \int_{0}^{s}ds' \frac{1}{(s-s')^{5/2}}.
\end{eqnarray}
Taking the limit of $b\rightarrow \infty $, Eqs. (~\ref{eq:icd_lim}) and (~\ref{eq:idd_lim}) have been obtained.

\section*{REFERENCES}
\setcounter {equation} {0}
\pagestyle{empty} \label{REFERENCES}

\newpage
\section*{FIGURE CAPTIONS}
\pagestyle{empty}

\begin{description}
\item[Fig. 1.:] Cartoon of a polyzwitterionic chain modelled as a continuous curve of length 
$Nl$, $l$ being the Kuhn segment length. An arc length variable $s$ is used to represent any point 
along the curve. The polyzwitterionic sites on the side groups in a real chain are modelled 
as dipoles (each with dipole moment of $\mathbf{p}_m$) oriented at some angle to the continuous curve.  
\end{description}

\begin{description}
\item[Fig. 2.:] Dependence of the asymmetry in the degree of counterion adsorption ($= |\alpha_{+} - \alpha_-|$) on the specificity of the salt is demonstrated in this figure. Figures (a), (b) and (c) correspond to $\delta_{+} = \delta_{-} = 3.5$; 
 $\delta_{+} = 3.5, \delta_{-} = 2.0$ and $\delta_{+} = 3.5, \delta_{-} = 7.0$, respectively. These figures 
 are obtained for $c_s = 1 mM, N = 1000, Nl^3/\Omega = 0.001, p_m/el = r_d/l = 1$, and $p_{\pm}/el = 0.1$.
\end{description}

\newpage
\vspace*{1.0cm}



\newpage

\end{document}